\documentclass{aa}
\usepackage{natbib}
\usepackage{graphicx}          
\usepackage[figuresright]{rotating}
\usepackage{ulem}
\usepackage{amsmath,amssymb}

\newcommand{\ms}{\mbox {$M_{\odot}$}}
\newcommand{\msun}{\mbox {$M_{\odot}$}}

\newcommand{\md}{\mbox {$\dot{M}$}}

\newcommand{\ls}{\mbox {$L_{\odot}$}}
\newcommand{\myr}{\mbox {~${\rm M_{\odot}~yr^{-1}}$}}
\newcommand{\ledd}{\mbox {$L_{\rm Edd}$}}

\newcommand{\te}{\mbox {~$T_{\rm eff}$}}
\def\apgt{\ {\raise-.5ex\hbox{$\buildrel>\over\sim$}}\ }
\def\aplt{\ {\raise-.5ex\hbox{$\buildrel<\over\sim$}}\ }

\begin{document}
\title{
On the evolution and fate of super-massive stars
}

\author{L. R. Yungelson\inst{1,2,3}
        \and
        E. P. J.  van den Heuvel\inst{1}
        \and 
        Jorick S. Vink \inst{5}
        \and
        S. F. Portegies Zwart\inst{1,4}
	\and
	A. de Koter\inst{1}
}

\offprints{L. Yungelson, lry@inasan.ru}

\institute{ Astronomical Institute ``Anton Pannekoek'', 
            Kruislaan 403, NL-1098 SJ Amsterdam, the Netherlands
           \and
           Institute of Astronomy of the Russian Academy of
            Sciences, 48 Pyatnitskaya Str., 119017 Moscow, Russia 
             \and
              Isaac Newton Institute, Moscow branch, 13
            Universitetskii  pr.,  Moscow, Russia
             \and
	     Section Computational Science, Kruislaan 403, NL-1098 SJ
            Amsterdam, the Netherlands
                \and
              Armagh Observatory, College Hill, BT61 9DG, Northern Ireland, UK
}
\date{received \today}

\titlerunning{Super-massive stars}

\authorrunning{Yungelson et al.}

\abstract {We study the evolution and fate of solar composition
super-massive stars in the mass range 60 --- 1000 \ms. 
Our study is relevant for very massive objects
observed in young stellar complexes as well as for super-massive stars that
could potentially form through runaway stellar collisions.}  
{We predict
the outcomes of stellar evolution by employing a mass-loss prescription that is
consistent with the observed Hertzsprung-Russell Diagram location of
the most massive stars.}  
{We compute a series of stellar models with
an appropriately modified version of the Eggleton
evolutionary code.}
{We find that super-massive stars with initial
masses up to 1000 \ms\ end their lives as objects less massive than
$\simeq 150$\,\ms. These objects are expected to collapse into black
holes (with $M \la 70 \ms$) or explode as pair-instability supernovae.}  
{We argue that if ultraluminous X-ray sources (ULXs) contain 
intermediate-mass black holes, these are unlikely to be the 
result of runaway stellar collisions in the cores of young clusters.}


\maketitle

\section{Introduction}
\label{sec:intro}

In this paper, we study the structure and evolution of 
very massive stars (VMS), defined as objects with masses of 60 up to 150\ms,
as well as super-massive stars (SMS) with masses in the range 150 -
1000 \msun.

The interest in the upper limit of stellar masses and the
evolution and fate of the most massive stars in the Universe was
greatly boosted by the discovery of ultraluminous X-ray sources
\citep[ULX, see][ and references
therein]{fabbiano89,colbert04,fabbiano03,fabbiano_x_sources06,soria_recipes06}. These
objects are most commonly interpreted as binaries involving either
sub-Eddington accretion onto an \textit{intermediate mass black hole}
(IMBH) with mass $\sim (10^2 - 10^5)$\,\ms\ or super-Eddington
accretion onto a \textit{stellar mass black hole} with mass
$\sim10$\,\ms.  In the latter case, beaming or support of
super-Eddington luminosity by an accretion disk is required.
Currently, the issue of the black hole masses in ULXs remains
unresolved \citep[see, e.g.,][]{fabbiano_x_sources06}.

It has been argued that IMBHs may also reside in the cores of
some globular clusters \citep[see,
e.g.,][]{gerssen02,gerssen03,gebhardt05,Patruno:2006bw,feng_ulx_imbh06}, but 
see \citet{baumgardt_m15_2003,baumgardt_g1_2003} for counter-arguments. 
The masses of the putative black holes in
the cores of well-known globular clusters such as 47~Tuc and NGC~6397,
are expected to be $\sim (10^2 - 10^3)$\,\ms\ \citep{DeRijcke:2006tu}.

ULXs are observed in both spiral and elliptical galaxies, i.e. in
environments with diverse metal abundances and star
formation rates. In this paper, we focus on stars with solar
 initial composition
($X$=0.7,$Z$=0.02). 
This choice is partly motivated by the high
metallicity of the starburst galaxy M82 -- with $Z\simeq Z_{\odot}$
\citep{mcleod_m82_solar93,origlia_m82_chem04,mayya_m82_chem06} -- 
which contains one of the most promising candidate ULXs, M82~X-1, and that may host a black hole 
of intermediate mass, as argued by
e.g. \citet{ptak_m82_imbh99,kaaret_m82_imbh01,matsumoto_m82_ulx01,strohmayer_m82_imbh03},
but see \citet{okajima_noimbh_m82} for arguments in favour of a stellar mass black hole in M82~X-1. 

The current observational estimate of the upper cut-off of stellar
masses is $\sim 150$\,\ms\
\citep{massey_hunter98,weidner_kroupa_upper_m04,figer_upper05}.
However, for solar chemical composition, it is expected that even such
massive stars rarely produce black holes more massive than $\simeq
20$\,\ms\ \citep{maeder92,fryer_kal_bh01}, 
due to copious  mass loss during both the
hydrogen and helium burning stages. If ULXs and young globular
clusters really harbour black holes with masses exceeding several
10\,\ms, the problem of their formation becomes a challenge for the
theory of stellar evolution with mass loss.

Related intriguing problems involve the formation and evolution of the 
 luminous stars found in the central parsec
of the Galaxy, e.g., the S-stars and the IRS13 and IRS16
conglomerates
\citep{schodel_irs13,lu_irc16,maillard_13e}\footnote{Note that a
high stellar mass inferred from luminosity may be an artifact of
unresolved binarity.  For instance, IRS16 SW turned out to be a binary
with two almost identical components of $\sim 50$\,\ms,
\citep{depoy_irs16,martins_irs16sw_bin06,peeples_irc16sw_bin06}.}, or
in the R136 stellar complex of the 30 Doradus nebula in the Large
Magellanic Cloud \citep{walborn97,massey_hunter98}.

Returning to the problem of IMBH formation, there are currently two
models in favour.  One involves the gradual accumulation of mass by
accretion onto a seed black hole, which, while swallowing gas and
stars in a stellar cluster, may grow to the intermediate mass range
\citep[see, e.g., ][ and references
therein]{miller_imbh_form04}. 
The other scenario considers the collapse of an
object which descends from an SMS formed by hierarchical runaway
merger of ordinary stars in a young dense stellar cluster, during or
after core collapse \citep[see, e.g.,
][]{pmmh99,pm_imbh_clust02,mbphm04,gfr04,pbhmm_clust04,fgr06}.

The motivation for the current study stems from the latter scenario.
It assumes that stars are born with a wide distribution of masses
ranging from the hydrogen-burning limit up to a maximum of about
150\,\msun.  More massive stars observed in the Galaxy may originate
from stellar mergers.  Most of these may be the result of coalescence
of components of binaries in common envelopes. 
The mass of binary
merger products may be up to 300\msun.
  \citet{pmmh99,miller_hamilton02,pbhmm_clust04,gfr04} have demonstrated by
means of detailed N-body simulations, in which simple stellar evolution was
taken into account, that there is a range of initial conditions where
stellar coagulation drives the mass of a star to $\sim1000$\,\msun.  
The resulting star burns-up quite quickly, and the process
of hierarchical merging in the cluster core terminates as soon as the 
first massive stars experience supernovae and collapse into black holes.   
The collision runaway process terminates as the mass loss
from the explosions of massive stars in the cluster center drives 
the expansion of the cluster core.  
This scenario was used to explain
the large black hole mass of M82 X-1 \citep{pbhmm_clust04}. 

The studies of hierarchical mergers take into account the
possibilities of collisional mass loss, mass loss from stellar winds,
and rejuvenation of merger products due to fresh hydrogen supply in
collisions.  However, stellar evolution is treated rather crudely in
these simulations, using extrapolations by several orders of magnitude
for stars that typically have zero-age main-sequence (ZAMS) mass $\leq
100$\,\ms. The possible formation of core-halo configurations and
the existence of an upper stellar mass limit are generally ignored. The
treatment of mass loss in stellar winds is particularly uncertain. In
principle, the winds of very massive and super-massive stars may be so
strong that the cluster core expands in such a dramatic way that it
stops being dominated by collisions, in which case the hierarchical
merger is terminated \citep{pmmh99,vanbev_dyn05}.

We aim to refine evolutionary calculations for merger products, and as a first step, we study 
the evolution of solar composition VMS and SMS over the mass range 60-1000\,$\ms$.  
We construct chemically homogeneous models for these stars and confirm the existence 
of their upper mass limit ($\simeq 1000$\,\ms), and study their evolution with mass loss 
through the core hydrogen and core helium-burning stages.  
This allows us to determine the mass and nature of the pre-supernova objects and to predict the 
fate of these objects.

The results of our evolutionary computations are presented and discussed in
Sect.~\ref{sec:evol}, a comparison with observations is given in
Sect.~\ref{sec:obs}, the fate of SMS is discussed in
Sect.~\ref{sec:fate}. Conclusions of our study follow in 
Sect.~\ref{sec:concl}. In the Appendix, we discuss a number of individual stars 
in close proximity to the Humphreys-Davidson (HD) limit.

\section{The stellar evolution calculations}
\label{sec:evol}

We compute the inner structure and evolution of single non-rotating
SMS.  Initially, all stars are chemically homogeneous. The products of
hierarchical merging are expected to grow in mass in discrete steps by
the injection of other stars and as a result the merger product may
be rapidly rotating.  In this paper, we ignore the effects associated
with merger events themselves, but concentrate on the 1D
(non-rotating) evolution of the final object, the super-massive star
(SMS).

Another open question, which has persisted for decades, is that of the 
pulsational stability of SMS \citep[see, e.g., ][]{baraffe_upper01}.  Existing
nonlinear calculations of the effects of pulsations on mass loss differ by an order of magnitude. 
\citet{appenzeller70} found that pulsationally-driven mass loss for 
solar metallicity stars with $M\geq300$\,\ms\ occurs on a timescale
that is shorter than the core-hydrogen burning timescale, but \citet{papa73}
 claimed that the mass-loss rate is an order of
magnitude lower and evolutionarily insignificant.
Wolf-Rayet (WR) stars and luminous blue variables (LBVs) are the descendants
of main-sequence stars and are likely to be subject to pulsational instabilities \citep{fadeyev04}.
Awaiting the resolution of this problem, we do not explicitly consider
the possibility of vibrational mass loss in our calculations.

The evolutionary computations were carried out by means of an appropriately
modified version of the \citet[][ priv. comm.  2003]{egg71}
evolutionary code.
The input physics has been described in \citet{pteh95}, but here we briefly list the sources of opacities and 
nuclear rates. The opacity tables are constructed using OPAL tables of \citet{igles_rogers96}, low-temperatures 
opacities are from \citet{alexander94}, whilst \citet{hubbard69} and \citet{itoh83} 
list conductivities for degenerate matter. 
The nuclear evolution of $^1H$, $^4He$, $^{12}C$, $^{14}Ne$,  $^{16}O$, $^{20}Ne$,  $^{24}Mg$ is followed using 
the reaction rates from \citet{caughlan88} and \citet{caughlan85}. 
The equation of state is based on the principle of Helmholtz free energy minimisation and 
provides the physical quantities such as the pressure, density, specific heats etc. as a function of a 
parameter related to the electron degeneracy and temperature.
\footnote{The most recent updates may be found at {\tt http://www.ast.cam.ac.uk/$\sim$stars/}}

An initial solar chemical composition ($X$=0.7 and $Z$=0.02) is assumed, and 
the computations may be relevant to the
Galactic objects such as the most massive members of Arches,
Quintuplet, IRS13, IRS16, NGC3603, Westerlund clusters \citep[see,
e.g.,][for further references]{PortegiesZwart:2004iw} and, in case
they do exist, to SMS in those external galaxies that have chemical
abundances comparable to the composition of the Milky Way.
We consider stars with surface helium abundance $Y_{\rm c} \geq 0.4$
and $T_{\rm eff}\geq 10000$\,K to be WR stars.

\subsection{Homogeneous supermassive stars}
\label{sec:bending}

With the opacity increasing outwards, a star develops a structure
where most of the mass is concentrated in the compact convective core,
whilst only a small fraction of the mass is located in a very extended
radiative envelope \citep[``core-halo configuration'',
][]{kato_85massive,kato_86sw}.  More massive stars tend to be more
extended, and the expansion of the envelope is caused by the increase
of the Eddington luminosity $\ledd = 4\pi c G M_{r} / \varkappa$,
where $\varkappa$\ is the flux-mean opacity in the radiative
outer layers of the star.  The maximum mass for a stable star is
reached when the luminosity of the star reaches \ledd\ in the
photosphere. Using Compton scattering opacity
$\displaystyle{\varkappa=0.2\,(1+X)/(1+2\times10^{-9}\,T)}$, Kato
succeeded in constructing homogeneous hydrogen-rich solar metallicity
stellar models of $10^7$\,\ms. The inclusion of the Kramers term in
the opacity resulted in a reduction of the limiting ZAMS mass to
$\approx 10^6$\,\ms\ \citep{mensch_tut_massive89}. These last authors
also found an upper limit of about 2500\,\ms\ for helium-rich stars
(X=0, Y=0.97).

\citet{ishiietal99} explored the core-halo effect using modern OPAL
opacities and found that, in this case, the upper mass limit 
for hydrogen-rich stars drops drastically.  For stars of solar
composition, this limit is reached at a mass of about 1000\,\ms, and
at a somewhat lower mass for stars with lower metallicity. 

\begin{figure*}
\includegraphics[angle=-90,scale=0.6]{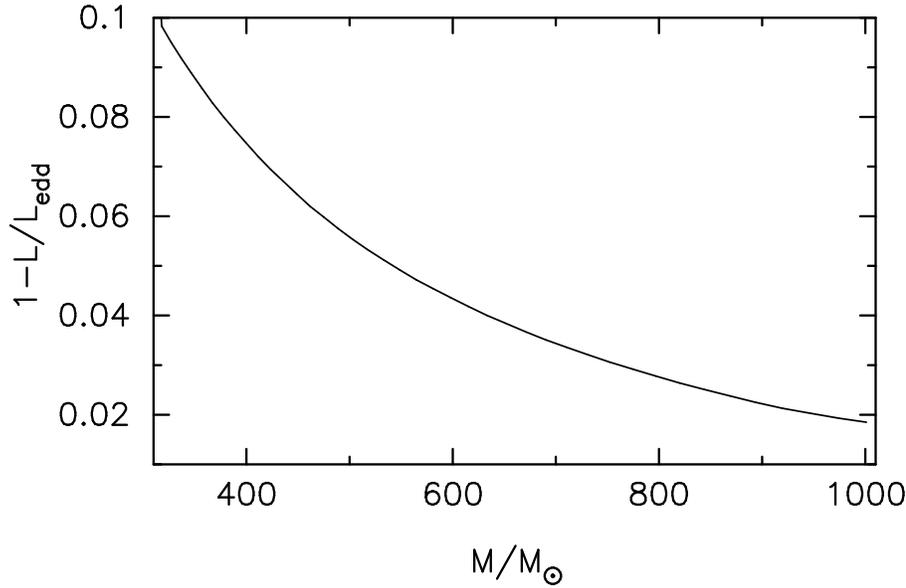}
\caption[]{The behaviour of the factor $\Gamma=1-L/\ledd$ in the outermost meshpoint of the models along the sequence of homogeneous models.
}
\label{fig:1ledd}
\end{figure*}

If $\displaystyle{\Gamma \equiv L/L_{\rm Edd} \rightarrow 1}$\ in the
interior of a 1-D stellar model, the radiative pressure can be balanced by a
density inversion \citep[see, e.g., ][]{langer97}.  In our
calculations, the highest luminosity is reached for homogeneous
models for $M \approx 1001$\,\ms. In this case, $\Gamma \approx 0.98$ 
in the outermost meshpoint in the stellar model (see Fig.~\ref{fig:1ledd}).  
A further increase of the stellar mass is then not possible, since
the convection in the surface layers is insufficient to transport the
stellar luminosity, and density inversions cannot build up.

If stars rotate and the critical velocity of rotation behaves as
$v_{\rm crit} \propto (1 - \Gamma)^{0.5}$, critical rotation is
achieved before $\Gamma$ reaches unity for any rotational velocity
\citep{langer97}.  For rotating stars, the actual upper mass
limit for solar metallicity may therefore be lower than 1000\,\ms.

The development of a core-halo structure leads to a bending
of the sequence of homogeneous stars towards the red in the HRD.  
In our models, this bending occurs above a mass \mbox{$\simeq 133$\,\ms}\ 
in agreement with e.g., \citet{figer_pistol98,ishiietal99,stothers_chin99}.

An approximate mass-luminosity relation for (100 -- 1000)\,\ms\
homogeneous stars obtained in the present study is (in solar units):
\begin{equation}
  L\simeq10^{3.48}M^{1.34}.
\label{eq:l-m}
\end{equation}
For the (25 -- 115)\,\ms\
initial mass range, the mass-radius relation may be
approximated as
\begin{equation}
R\simeq10^{-0.9}M^{1.02},
\end{equation}
and for larger masses as
\begin{equation}
  R\simeq 10^{-0.77} M^{0.96}
\end{equation}

We note that although we can theoretically construct this sequence of completely homogeneous massive star models,
  in reality, stars may form by accretion, and stars could start core hydrogen burning before accretion halts, and evolve
 along the 
main-sequence until the accretion reservoir becomes exhausted \citep[see, e.g., ][]{behrend01}.
  
\subsection{The mass-loss rate}
\label{sec:mdot}

Massive stars are subject to considerable mass loss, driven mainly by
radiation pressure through spectral lines, which can be enhanced by
currently poorly understood vibrational modes. For stars more massive
than $\sim 100$\,\msun\, theoretical models are not yet well
developed. \citet{kudr02} calculated mass-loss rates for masses up to
300\ms, but the mass-loss rates of
Kudritzki do not include the important effect of multi-line
scattering, which may be vital for models in close proximity to the
Eddington limit.  \citet{vink06} presented mass-loss predictions for
very massive stars close to the Eddington limit, and found a steep
behaviour of the mass-loss rate as a function of $\Gamma$ (but for a
constant ratio of the terminal velocity over the escape velocity).

We did not use the mass-loss rates of \citet{vkl00} for stars above 
 100\,\ms, as they were derived for $\Gamma < 0.5$. 
As the Vink et al. recipe includes 
bi-stability jumps, which are a function of $T_{\rm eff}$ and wind density (and hence $\Gamma$), we did 
not extrapolate the Vink et al. recipe outside its validity range.

A break of evolutionary sequences occurred in calculations for massive
stars by other authors with mass-loss prescriptions based on radiation-driven
wind models \citep[e.g., for Pistol star by][]{figer_pistol98}. The
extrapolation of fits to the empirical data on mass-loss
rates \citep{dnv88} for stars more massive than $\simeq120$\,\ms\,
results in unreasonably high mass-loss rates, causing the stellar
evolution models to lose their ability to converge.

Due to the sparsity of theoretical expressions for the amount of mass
loss from supermassive stars, we adopt the following \textit{ad-hoc}
``momentum'' equation for our evolutionary models:
\begin{equation}
\label{eq:mdot}
\dot{M} = \frac{L}{v_\infty\,c} \frac{1}{(1-\Gamma)^{(\alpha-0.5)}},
\end{equation}
where $L$ is the stellar luminosity, $v_\infty$ is terminal
velocity of stellar wind, $c$ is the speed of light, and $\Gamma = L/L_{\rm
Edd}$ in the outermost meshpoint of the stellar model. After
\citet{kudr_puls00} we define 
\begin{equation}
  v_\infty=C(T_{\rm eff})\left[(2GM/R)(1-\Gamma)\right]^{0.5}.  
\end{equation}

We do not specify a particular physical mechanism 
for our mass-loss prescriptions, but the luminosity dependence in Eq.~(\ref{eq:mdot}) accounts 
for the assumption that radiation pressure almost certainly plays a role. Furthermore, we account 
for the proximity to the Eddington limit 
by introducing the $L/L_{\rm Edd}$ dependence. We have tested our Eq.~(\ref{eq:mdot}) against 
the \citet{vkl00} mass-loss recipe for a 60 \,\ms\ star, and although there are differences 
due to the presence of bi-stability jumps in the Vink et al. recipe, which we do not account for in 
our mass-loss prescription, the overall mass lost (35\,\ms) is 
very similar, bringing a 60 \,\ms\ star down to 25 \,\ms.

As noted by \citet{vkl01}, while the metal lines are responsible for driving 
the wind, hydrogen and helium lines are sparsely populated in the spectral 
range where early-type stars emit
most of their energy. This provides some justification for using the
same mass-loss recipe for hydrogen-rich stars, as well as for stars with
hydrogen-exhausted atmospheres. 

We base our choice of $\alpha$ in Eq.~(4) on two criteria: 1) stars should spend
most of their lifetime between the ZAMS and the
HD-limit (as defined in Humphreys \& Davidson 1994), and 2) they should spend
only a limited amount of time, e.g. less than $\sim 1 - 2$\% of their
lifetime, at effective temperatures cooler than the Humphreys-Davidson
limit.  The nature of the Humphreys-Davidson limit is still
elusive, as it is still unclear whether the lack of stars above this
empirical limit is due to physical processes (likely to be related to the
Eddington limit), or it has a statistical nature instead. As some
stars are observed on the cool side of the HD-limit (see
Fig.~\ref{fig:hrd_obs_col025}), we consider the possibility that
the evolutionary tracks extend to the yellow and red regions of
HR-diagram.

\begin{figure*}
\includegraphics[angle=-90,scale=0.6]{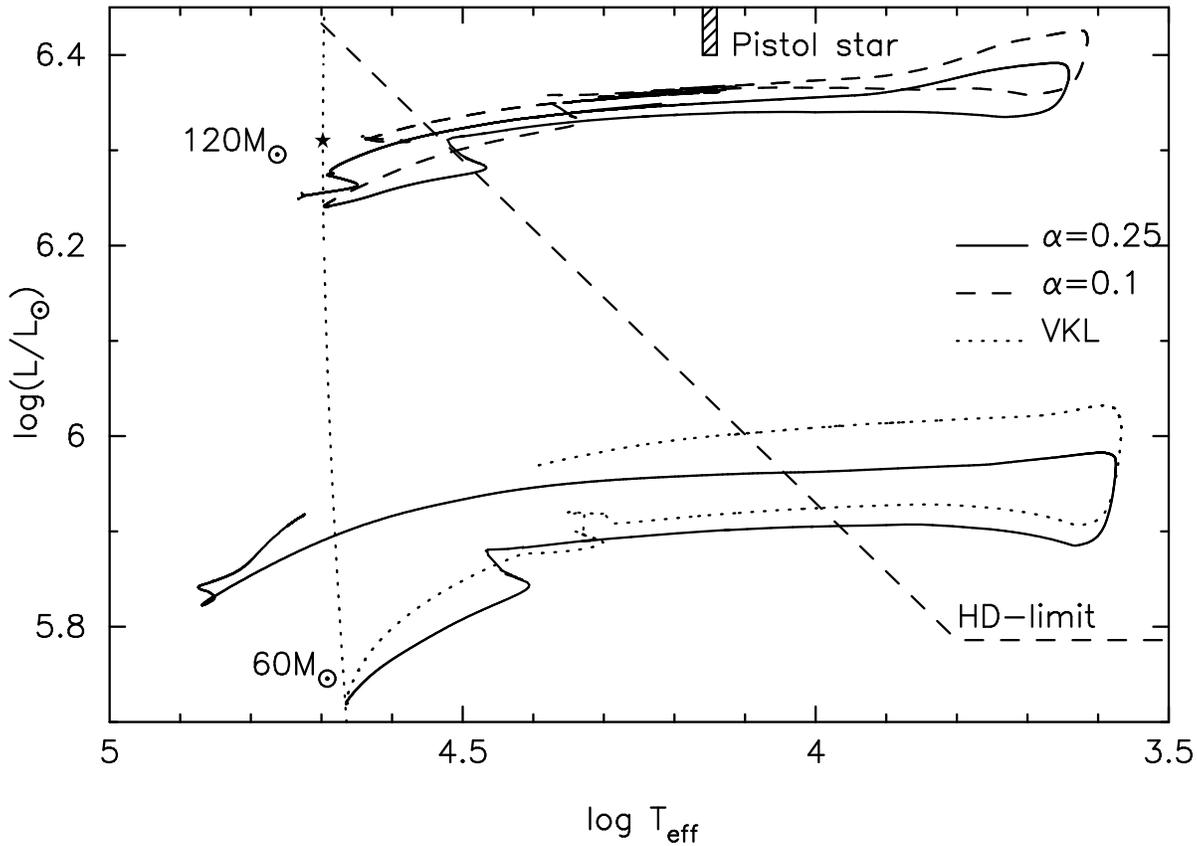}
\caption[]{Evolutionary tracks of stars with $M_0$=60 and 120\ms\ in the Hertzsprung-Russel diagram for different assumptions of the mass-loss prescription.
For 60\,\ms, the tracks for  mass-loss prescription given by Eq. (4) with $\alpha$=0.25 and mass-loss rates from 
\citet{vkl00} are shown. For 120\,\ms, tracks 
for Eq. (4) prescription with $\alpha$=0.25 and  0.1 are shown.
The long-dashed broken line shows the 
Humphreys-Davidson limit. The star symbol marks the bending of the sequence of homogeneous models.}
\label{fig:hrd_dj_025_01}
\end{figure*}

We present our simulations for 60 and 120\,\ms\ stars in 
Fig.~\ref{fig:hrd_dj_025_01} and Table~\ref{tab:tracks}. 

For $\alpha=0.25$, the $M_0=60\,\ms$ star loses 23.5\,\ms\ during the core-hydrogen
burning phase that lasts for 3.8 Myr. At the TAMS, the surface hydrogen and helium abundances 
are virtually equal $X\approx$0.49 and $Y\approx$0.49, and $C/N$ and $O/N$ ratios are very close to 
their nuclear equilibrium values (see Fig.~\ref{fig:chem} below).
The star may be classified as a strongly helium-enriched O-star.
A further 12\,\ms\ are lost in the part of the track where the stellar 
effective temperature is below $\simeq$ 10\,000K, with a lifetime in this later
stage of 0.048 Myr. When the star crosses the HD-limit moving blueward, the
hydrogen abundance at its surface is 0.096, and the object has become a WR star.

This track agrees well with the track for which we used the mass-loss prescription of 
\citet{vkl00}, where the mass-loss behaviour is not completely continuous due to 
the presence of bi-stability jumps. The total amount of mass lost during the main-sequence phase 
is 19.4\,\ms, whilst 13\,\ms\ is lost when star is cooler than $\sim$ 10\,000K. The lifetime of
the later stage is 0.088 Myr.  This compares well to the results of \citet{limongi_chieffi06}, who used
the mass-loss rates from \citet{vkl00} and found a total amount of mass loss of $\Delta M=22\,\ms$ during 
the main-sequence stage.

The $M_0= 120$\,\ms\ star, computed with $\alpha=0.25$,  
 loses $\simeq 57$\,\ms\ over its 
2.76 Myr long main-sequence lifetime.
While on the cool side of the HD-limit (at $T_{\rm eff} \aplt
30\,000$K), the star loses an additional $\simeq 14$\,\ms\ in 0.039
Myr. In the latter case, the star becomes a WR object during
the main-sequence (see more below).

The last models of the sequences for stars with initial masses 60  and 120\,\ms\
are in the core He exhaustion phase (Table~\ref{tab:tracks}), with masses  of
$\simeq 40$\,\ms\ and $\simeq 21$\,\ms, respectively. These values are in
reasonable agreement with the masses of $\simeq 30$\,\ms\ and $\simeq 20$\,\ms,
obtained by \citet{limongi_chieffi06} who used mass-loss rates from 
\citet{vkl00} for blue supergiants, from \citet{dnv88} for red supergiants,
and from \citet{nl00} for WR stars. We also note that the tracks computed by
\citet{limongi_chieffi06} for 60 and 120\,\ms\ stars penetrate into the cool
region of the HR-diagram to $\log T_{\rm eff} \approx$ 3.65 and 3.70,
respectively, again in good agreement with our computations. 

Finally, our tracks for $\alpha=0.25$ are in reasonable agreement with
the tracks computed using the spline fits of \citet{dnv88} for empirical
\md\, but, as we mentioned above, the latter fits diverge at high
stellar masses. 

For $\alpha=0.1$, the $M_0=120$\,\ms\ star
crosses the Humphreys-Davidson limit during core-hydrogen
burning, and spends $\sim$ 0.5\,Myr in the low-temperature region,
which is too long.

Based on this analysis, we conclude that $\alpha
\simeq 0.25$ is suitable for evolving supermassive stars
 and we adopt this value throughout 
 the paper.

\subsection{Evolution of supermassive stars}
\label{sec:comput}

\begin{figure*}
\includegraphics[angle=-90,scale=0.6]{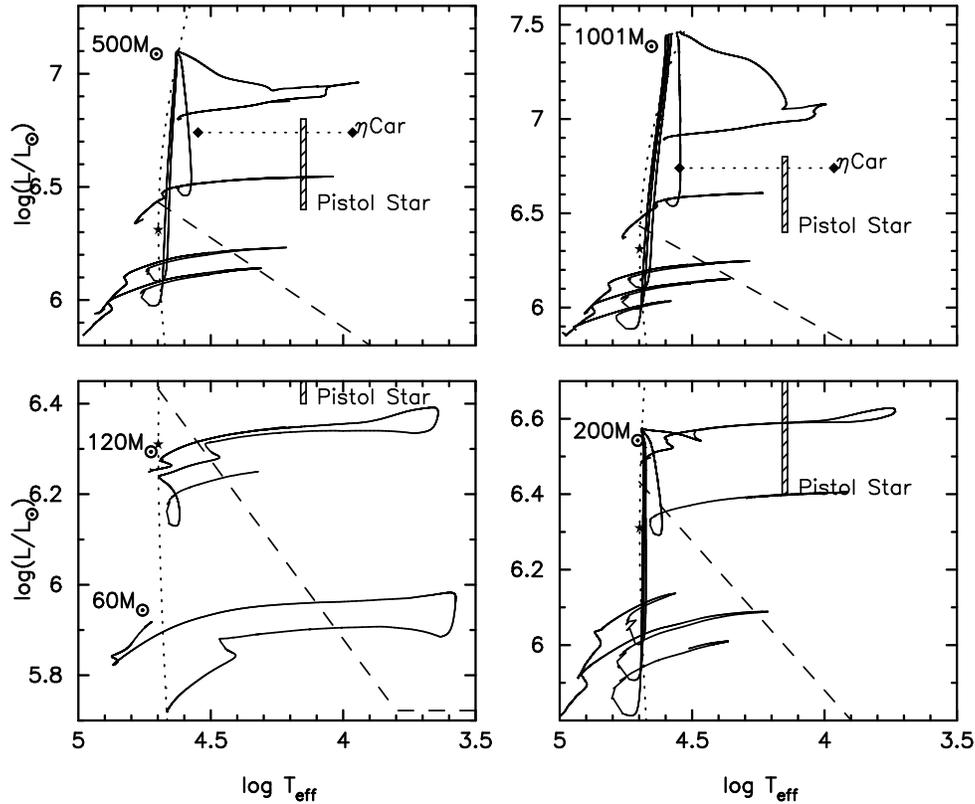}
\caption[]{Evolutionary tracks of stars in the Hertzsprung-Russel diagram.
For 60\,\ms, the track for $\alpha$=0.25 is shown; for 120\,\ms, tracks for $\alpha$=0.25 and 0,5 are shown; for 200, 500 and 1001\,\ms, tracks for 
$\alpha$=0.25, 0,5, 0.8, 0.9 and 1.0 are shown. 
The dotted line to the left shows the locus of homogeneous models.
The dashed line indicates the Humphreys-Davidson limit (extrapolated for high $T_{\rm eff}$).
The star symbol marks the position of a 133\,\ms\ star
 (indicating the bending point of the ZAMS).
For orientation,  the hatched rectangle shows the position of the Pistol star for the ``low-luminosity'' solution 
of \citet{figer_pistol98} and the diamonds show the position of $\eta$\,Car, which is somewhat uncertain because of the ambiguity in 
the radius determination \citep{hillier01}.  
 }
\label{fig:hrd_all}
\end{figure*}

A summary of the parameters of evolutionary tracks computed in the
present paper is given in Table~\ref{tab:tracks}.  
Figure~\ref{fig:hrd_all} shows evolutionary tracks for stars
with initial masses 60, 120, 200, 500, and 1001\,\ms, computed for
different values of the parameter $\alpha$. 

\begin{figure*}
\includegraphics[angle=-90,scale=0.6]{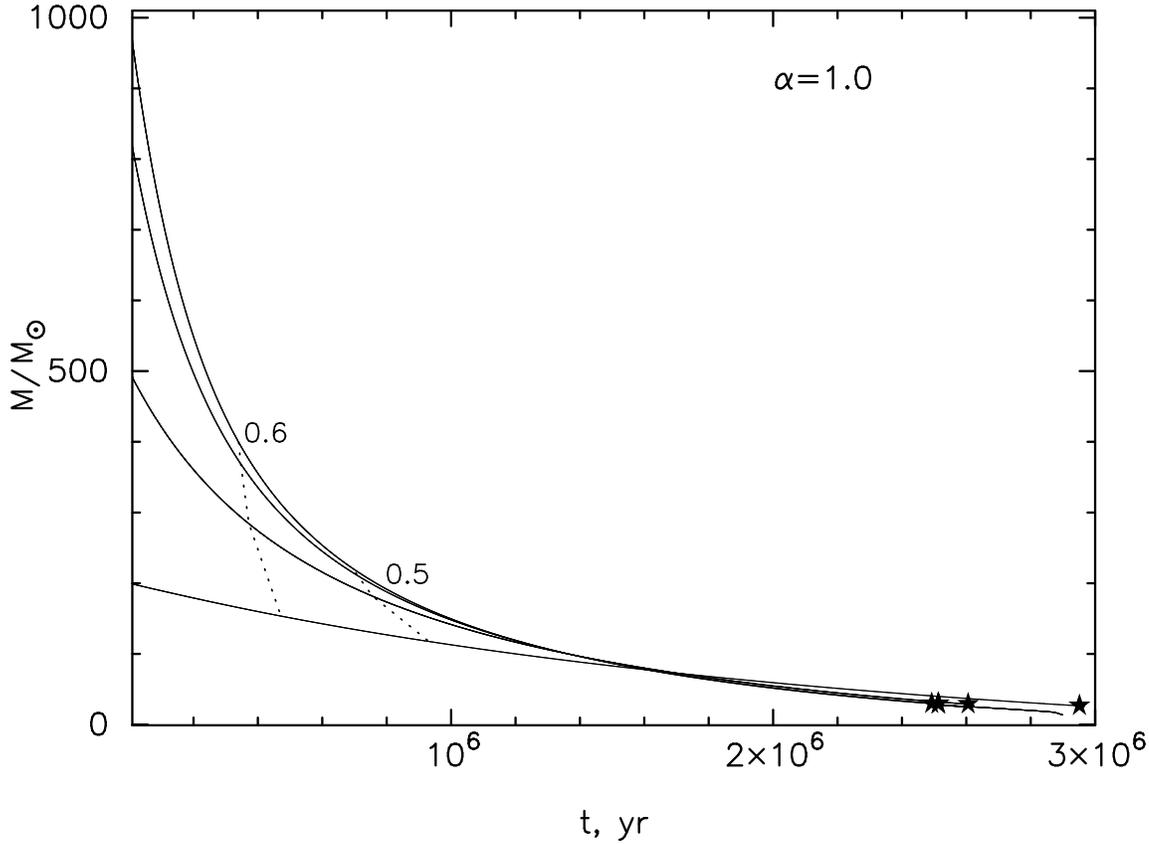}
\caption[]{Stellar mass vs. evolutionary lifetime for $\alpha$=1.0. Initial mass of stars is 1001, 843, 500, and 200\ms.
The dashed lines connect the loci of models with a hydrogen abundance in the convective cores of ${\rm X}_{\rm c}=$0.6 and 0.5.  
The star symbols at the curves indicate the central hydrogen exhaustion time (see also Table~1).
}
\label{fig:tm1}
\end{figure*}

For $\alpha \geq 0.5$, the mass-loss timescale is shorter than the
nuclear evolution timescale. The stars evolve downward along the 
locus of homogeneous stars remaining practically unevolved 
(Figs. \ref{fig:hrd_all},\ref{fig:tm1}).
For all sequences, the mass-loss rate is so high that
nuclearly-processed layers are almost immediately exposed at the
surface. For stars with $M_0$=500 and 1001\,\ms, when central hydrogen
becomes exhausted, the surface abundance of hydrogen is just several
per cent.  Thus, SMS turn into WR stars during the core hydrogen
burning stage. 

For models with high $\alpha$, the bending of the evolutionary tracks
to higher \te\ (bottom loop) starts when $X_c \simeq 0.05$ in the
convective core, i.e. this corresponds to what, in the common notation,
would be ``point B'' of the evolutionary track.  The blue points of
the loops of the tracks for $\alpha \geq 0.5$ correspond to the core
hydrogen exhaustion. The direction of evolutionary tracks changes from
redward to blueward, when the core He-burning becomes the dominant
source of energy: $L_{He}/L_{H} \sim 3$. For all models, the stars evolve into virtually
identical helium stars, as the stellar evolution converges.  
The evolutionary tracks start to deflect from the ZAMS to the right only for 
$\alpha \leq 0.5$.
These high-$\alpha$ tracks
may be relevant for the evolution of putative SMS ($M \apgt
120$\,\msun), but they are less relevant for known VMS ($M \aplt
120$\,\msun).

The behaviour of the evolutionary
sequences with high mass-loss rates is consistent with the results of 
\citet{maeder_massive80}, who showed that
stars with an initial mass in excess of 100\,\ms, and with a time
averaged $\md \apgt 2 \times 10^{-5}$\,\myr\ evolve towards the left
of the HR diagram.  In the interior, a large homogeneous core is
surrounded by an envelope with very little difference in composition
between the core and the stellar surface.  \citet{stothers_chin99} used OPAL opacities 
and found a similar convergence of the evolutionary tracks.

To be consistent with restrictions upon the mass-loss rates found 
by evolutionary computations for $M_0=$60 and 120\,\ms\ stars (\S~\ref{sec:mdot}), we now 
discuss the tracks for $\alpha=0.25$. The behaviour of all evolutionary sequences 
is rather similar. 
For high $\alpha$ ($\alpha$=1; Fig.~\ref{fig:tm1}) the tracks converge quicker
than for our favourite  $\alpha$=0.25 models (Fig.~\ref{fig:tmhy025}).
Figure
\ref{fig:tmhy025} represents the relations
between evolutionary time, the mass of the star, the central hydrogen
abundance and the surface helium abundance for $\alpha=0.25$.  For the
$M_0$=200 and 120\,\ms\ models, the hydrogen surface abundances for
${\mathrm X_c}=0$ are 0.12 and 0.25, respectively. If we
consider the surface helium abundance, $Y\approx0.4$,
as a conventional threshold for assigning the WR spectral type, this limit is
reached within $\lesssim$\,1Myr for stars with $M\geq 200$\,\ms,
whilst it takes about 2\,Myr for the 120\,\ms\ star.

 Figures~\ref{fig:tdmdt025} and \ref{fig:lledd_025} 
  show that after
the end of the central H-burning all stars start to expand on the
thermal time scale of their envelopes.  Expansion is accompanied by a
strong reduction of $L/L_{\rm Edd}$: although the stellar $L$ remains
practically the same, the $L_{\rm Edd}$ of the outermost meshpoint
increases rapidly due to a reduction in opacity. The reduction of
$\Gamma=L/L_{\rm Edd}$ would increase the mass-loss rate if it were
to act alone.                                                                                                                                                   
However, our adopted mass-loss prescription 
depends weakly on $\Gamma$. The decisive factor in Eq.~(\ref{eq:mdot}) is $v_\infty$, 
which decreases strongly with the expansion of the star, causing the growth 
of $\dot{M}$.
A further mass and gravity reduction at almost constant luminosity
results in a kind of runaway mass loss, i.e., the star becomes
unstable. The expansion terminates when the core He-burning becomes
the dominant source of stellar luminosity. The star then returns
rapidly to the blue region of the HR-diagram. In our calculations mass
loss occurs continuously.
In real stars, one may expect
that instability associated with high $L_{\rm Edd}$ would manifest
itself as a sequence of sporadic mass-loss episodes which terminate
when He-burning causes an overall contraction of the star.  We may
speculate that the total amount of mass lost in the ``spike'' of the
mass-loss rate during redward excursions is similar to that which
would be lost by a ``real'' star in a series of outbursts.  The total
stellar luminosity becomes very close to the Eddington luminosity when
the star returns to the high \te\ region of the HR-diagram.  

An interesting feature of SMS
evolution is the increase of the core hydrogen-burning time of the stars 
with increasing $\alpha$ in the mass-loss prescription.  
This is understandable, as heavy mass loss is accompanied
by a reduction of the central temperature and density (like in close
binary components that experience dynamical- or thermal-timescale mass
loss upon Roche lobe overflow).

Figure~\ref{fig:te_dmdt_025} shows that irrespective of initial
stellar mass, for $\alpha = 0.25$, the maximum mass-loss rates do not
exceed the maximum mass-loss rate for line-driven winds. 
\citet{smith_owocki_max_dmdt06} recently estimated this maximum to be:
\begin{equation}
\label{eq:dmdt_max}
\md \approx 1.4\times10^{-4} (L/(10^6\,\ls))\ \myr.
\end{equation}
A similar order of magnitude ($10^{-3}$\,\myr) estimate for the upper
limit of \md\ due to radiation pressure on spectral lines was obtained
by \citet{aerts04}.  
Figure~\ref{fig:perfn} shows the run of the
``wind performance number'' $\eta= (\md v_\infty)/(L/c)$, along
evolutionary tracks for model sequences with $\alpha=0.25$.

One of the main results in the section is the finding of the convergence of our 
models with a wide range of initial masses (see Fig.~\ref{fig:tmhy025}).
This convergence of stellar masses which initially differ by a factor $\sim$10 
may crudely be explained as follows. 
During most of the stellar lifetime the value of $L/\ledd$ is not far from 1 (see Fig.~\ref{fig:lledd_025}), whilst the 
stellar radius changes significantly for only a very short time. 
Therefore, the crucial factor that sets the rate of mass loss is the stellar luminosity, which may, as a first approximation, be 
taken as the luminosity of a homogeneous star.
For the case of $\alpha=0.25$, the stellar lifetime is $t \approx 10^{6.9}(M/\ms)^{-0.2}$\,yr (for $120\,\ms \leq M_0 \leq 1000\,\ms$). 
When we now apply the mass-luminosity relation, the amount of mass lost, $\Delta M$, is found to be 
comparable to the initial mass of the star, $M_0$.

\section{Models of VMS versus observed stars}
\label{sec:obs}

Figure~\ref{fig:tdmdt025} shows the mass-loss dependence on time for
$\alpha=0.25$\ sequences and compares them with the mass-loss ranges found
for young massive stars in the Arches and Quintuplet clusters, and the
cluster of HeI-emission stars in the Galactic center,
as well as for a sample of WNLh stars in young massive clusters in the Milky Way and the Magellanic Clouds (A. de Koter, unpublished). 
In Figure \ref{fig:mmdot}, we show in detail the relations between stellar mass and the mass-loss rate for $\alpha=0.25$ evolutionary 
sequences for objects with initial masses of 60, 120, and 200\,\ms, and we compare them with mass-loss prescriptions suggested 
for massive stars, as well as with observations of  known objects. We remind the reader that we consider stars with surface 
helium abundance $Y_{\rm c} \geq 0.4$ and $T_{\rm eff}\geq 10000$\,K to be WR stars.

A fit for observed mass-loss rates of O-stars is given by \citet{dnv88}. For $M > 120$\,\ms, 
comparing models  with any observational data fits is
not very sensible, because  it would involve too rough an extrapolation. 
However, our mass-loss rates for the $M_0=120$\,\ms\ star, when it is still
H-rich, do not contradict the fits of \citet{dnv88}, nor the observations of the
most massive H-rich WR-stars.  

For WR-stars, fits to observational data were given by e.g.
\citet{lan89b}\footnote{The numerical coefficient in Langer's formula 
 is taken to be three times lower than the originally suggested lowest coefficient, to account for wind 
clumping \citep{hamann_koest_wr98,marchenko_wr_clump07}.}: 
$$\displaystyle{\dot{M}=-2 \times 10^{-8} M^{2.5}};$$
\citet{nelemans01}: $$\displaystyle{\dot{M}=-1.4 \times 10^{-8} M^{2.87}};$$
\citet{donder_vanbev_wr03}\footnote{For the most massive WR stars, this fit gives rates close to
the ones suggested by the \citet{nl00} fits. In Fig.~\ref{fig:mmdot}, the original
formula is extrapolated to higher stellar masses.}:
$$\displaystyle{\log(-\dot{M})=\log L -10 }$$
(solar units and rates per yr are used).
In Fig.~\ref{fig:mmdot}, we also show observational estimates for mass-loss rates of WN stars 
with $M \geq 20$\,\ms\ from 
\citet{hamann_wn06}. Estimates of \md\ for  WR stars \citep{hamann_wn06} shown in Fig.~\ref{fig:mmdot} agree well with the estimates 
of \citet{cappa_wr04}, based on radio-observations that showed that $\dot{M}_{\rm WR}$ hardly exceeds $10^{-4}$\,\myr.  
Figure~\ref{fig:mmdot} clearly shows that extrapolation of the formulae given above to
very massive WR stars is not justified. 
Late-type WN stars like WR20a (see Appendix for more details) may still be in their core-hydrogen burning stage.

In Figs.~\ref{fig:hrd_obs_col025} and ~\ref{fig:hrd_obs_col05}, we
present evolutionary tracks for $\alpha$=0.25 and 0.5 sequences in the
HR-diagram and we compare them to the positions of some of the most
luminous stars in the Milky Way and the Magellanic Clouds
(Table~\ref{tab:obs}). 
For notes on  the individual stars shown in
Figs.~\ref{fig:hrd_obs_col025} and~\ref{fig:hrd_obs_col05}, we refer 
the reader to the Appendix.
These figures indicate quite clearly that $\alpha = 0.5$\ would be too high a value, and that 
$\alpha$ should be $\sim$ 0.25 to be consistent with the observations.

The data presented in Figs.~\ref{fig:tdmdt025},~\ref{fig:mmdot},~\ref{fig:hrd_obs_col025},
and~\ref{fig:hrd_obs_col05} suggest that the initial masses of the stars observed
in the Arches cluster, Quintuplet and R136 may exceed $\simeq
100$\,\ms.  Shortly after leaving the ZAMS, these stars acquire
surface abundances that result in their spectral classification as
transitional types between O-stars and WR stars.  The positions of
luminous stars with estimated $M$ and \md\ in Figs.~\ref{fig:mmdot}
and \ref{fig:hrd_obs_col025} suggest that they have initial masses up
to $\sim 100$\,\ms, and are in the core-hydrogen burning stage.
The positions of these objects in the $(M-\md)$-plots and the
HR-diagram are consistent.  This point may be further explored when
more detailed grids of evolutionary tracks become available.  We note
that masses and the estimates for \md\ are both uncertain.  For instance,
\citet{repolust04} estimate the range of uncertainty of their
spectroscopic mass determinations for Galactic O-stars (listed in
Table~\ref{tab:obs}) as $\displaystyle{_{-30}^{+50}}$\%. For the
mass-loss rates, the errors are up to 0.2dex.

The mass-loss rates that we obtained for pure helium stars have the
same range as \md\ for hydrogen-rich stars.
This is consistent with the similarity of \md\ values found for O-stars and HeI-emission 
stars, of which some are helium rich\footnote{The estimates of \md\ for HeI stars strongly 
depend on an uncertain He/H-ratio.}, but it does not seem to agree with the 
\citet{hamann_wn06} data for hydrogen-poor WR stars.

We should note that we obtain higher masses for WR stars  and pre-supernovae masses than in 
some previous computations for non-rotating stars. For instance, for the same 
120\,\ms\ star, \citet{schaller92} obtain $\simeq 7.6$\,\ms\ at the end of their 
calculations, whilst our computations yield a final mass of 40.4\,\ms\ (for $\alpha=0.25$).  
The difference stems from the difference in mass-loss rates for He-rich stars. For the latter, 
Schaller et al. use the rates from \citet{lan89b}, uncorrected for clumping, that are much higher 
than those given by our empirical algorithm for \md.   

In Fig.~\ref{fig:chem}, we show the evolution of the chemical abundances at the surface 
of a 60 \msun\ VMS and a 500 \msun\ SMS. We note several differences in the behaviour of 
the surface abundances. On the surface of the $M_0=$60\,\ms\ star, helium starts to 
dominate over hydrogen only close to the TAMS, whilst for the $M_0=$500\,\ms\ SMS, this 
already happens much earlier. 
This difference is due to the relative sizes of the convective cores. 
Furthermore, the surface of the VMS becomes enriched in nitrogen (relative to the solar abundance) 
close to the TAMS, and remains so until the end of the helium exhaustion of the core. 
For this case, we expect the pre-supernova object to be a WN star.

The surface layers of the SMS become enriched in nitrogen almost immediately after the departure from the 
ZAMS. Later, at the TAMS\footnote{With the caveat that we use the definition of the 
TAMS in a broad sense indicating the end of the core hydrogen-burning stage.}, the hydrogen and nitrogen 
abundances drop almost simultaneously, and the surface becomes dominated by He and C. 
In the final stages of evolution, oxygen becomes the dominant surface element, whilst
the abundance of helium becomes even lower than that of carbon. In this case, the expected pre-supernova object 
is an oxygen-carbon star, with traces of helium at the surface.

Figure~\ref{fig:chem} also shows the differences in the surface CNO-cycle 
element ratios of VMS and SMS. For the latter, the layers in which the elements are in nuclear 
equilibrium are exposed virtually immediately at the beginning of their evolution (after the loss of only 
about 100\,\ms\ in only $\simeq$0.7 Myr). This situation persists until C and O enriched layers 
become exposed due to He-burning. As the core H-burning evolutionary stage is by far the longest, we conclude 
that a majority of SMS should have nuclear-equilibrium ratios of CNO-elements. 

For the case of the VMS, the transition in chemical abundance from solar to nuclear-equilibrium 
occurs somewhat more smoothly. It starts after about 2 Myr of evolution, whilst it takes about 
2 Myrs out of a total 6 Myrs MS-lifetime. During this transitional phase, the star loses 
about 10\,\ms. If our mass-loss algorithm is applicable to the lower mass VMS, we should expect 
this transition to occur smoothly. VMS close to the TAMS are expected to be associated with 
luminous blue variables (LBVs), for which a wide span of CNO-elements ratios 
has been found. 
Although \citet{smith_lbv_chem98} found LBV ejecta to be $N$-enriched, these have 
generally not yet reached CNO equilibrium values. 
However, as there is controversy about the origin of LBV nebulae, it would arguably be better, or at least more direct, 
to consider photospheric abundances instead. These, however, may depend on complexities of the atomic physics and wind analyses, and 
the results are conflicting. For instance, evidence for advanced nuclear processing was 
found by \citet{lennon_lbv_chem03} for the LBV R71 in the LMC, but this is not a well-established fact for LBVs as a population.
Although most LBVs show 
CNO-enriched atmospheres, it seems unlikely that all of them have already reached their CNO equilibrium values. 

A range of CNO abundances may naturally be explained from our VMS computations, but given 
that most LBV masses are not significantly over 150\msun, it seems unlikely
that all LBVs are the progeny of SMS, as our models would predict SMS
to reach CNO equilibrium almost instantaneously.

\section{The fate of supermassive stars}
\label{sec:fate}

According to our evolutionary calculations, SMS transform into 
oxygen-neon stars with traces of carbon. Depending on the progenitor mass, 
the final masses of the stars are between
$\simeq$20\,\ms\ and $\simeq$140\,\ms.  One may expect that further
reduction of the total stellar mass would be negligible, because of
the short lifetime of stars in their post core-helium burning
evolutionary stages.  The helium-carbon-oxygen outer layer comprises only
a few per cent of the total mass of the star.  Thus, one may expect
that if these objects were to explode, SMS would probably produce
type Ic events.

It has been known for decades \citep[see, e.g., pioneering papers of
][and many other studies]{rakavy1967,fraley1968,ober_pairs83} that if
the mass of oxygen stars exceeds several tens of \ms, during central
oxygen-burning, they enter the electron-positron pair instability
regime, and contract quasi-dynamically.  When the central temperature
increases to $(3-6) \times 10^9$\, K, central oxygen burning becomes
explosive, which is much faster than neutrino energy losses. The
released nuclear energy may be sufficient for the internal energy to
exceed the gravitational binding energy. The star will then disrupt
completely, without leaving a compact remnant, giving rise to a
so-called ``pair-instability (or pair-creation) supernova'' (PISN).
If the released energy is not large enough to disrupt the star,
the star collapses to a black hole.  This instability arises
irrespective of the metallicity of the progenitor star.  However, for
solar composition stars the possibility of PISN is usually not
considered, since in the evolutionary models for even the most massive
stars formed in the ``standard'' way (i.e. without possible runaway
collisions), with masses up to ``observational'' upper limit of
150\ms, sufficiently massive oxygen cores are not formed.  We note
that this may stem from an overestimate in the mass-loss rates
during the WR-stage (if mass-loss prescriptions like those shown in
Fig.~\ref{fig:mmdot} are extrapolated).
   
However, massive oxygen cores are more often considered in relation to the evolution 
of Population~III stars. Massive oxygen cores may experience only
moderate mass loss, owing to the absence of metals (but see Vink \& de Koter 2005 and Vink 2006 
for a discussion on the possibility of higher mass loss due to the radiative driving through 
intermediate mass CNO elements, and the likely proximity to the Omega/Eddington limit). 
One may use the results obtained for Pop.~III~stars for solar composition stars,
because of a rather modest difference in internal structure of massive
stars of different initial metallicity during the late stages of
evolution \citep[e. g., ][]{schaller92}.  \citet{umeda_nomoto02} find,
in computations for Pop.~III stars, that PISNs are experienced by stars
that form He-cores with mass (70 -- 129)\,\ms.  A similar range of (63
-- 133)\,\ms\ was found by \citet{heger_woosley_pairs02} from
computations of non-rotating He-stars.  
Apart from differences in the
input physics, some discrepancy in mass obtained by different authors
stems from the well-known 
fact that the evolution of initially naked helium stars is slightly different from the evolution of similar mass helium cores 
of stars formed through hydrogen burning.  
Stars that have helium cores more massive than the ranges of PISN-progenitors quoted above 
form black holes without ejecting matter. Stars with He-core masses below these limits are 
expected to collapse directly to a black hole or form a black hole through fall-back \citep{fry99}.

In Fig.~\ref{fig:mimf}, we plot the initial-final mass relation for
SMS, using the results of
\citet{umeda_nomoto02} and \citet{heger_woosley_pairs02}, and we mark approximate
boundaries between regions with different outcomes.  These boundaries
are rather crude, but they suggest the following: we hardly expect the
formation of black holes with masses larger than $\sim$ 150\,\ms.  As
the difference between the upper mass boundary of PISN-producing SMS
($\sim$ 800\,\ms) and the upper SMS initial mass limit ($\sim$
1000\,\ms) is quite small, black holes with masses larger than 
PISN-producing objects may hardly form, and it is quite likely that the
most massive black holes produced by SMS have only $M \aplt
70$\,\ms. 

Figure~\ref{fig:life} shows the lifetimes of supermassive stars. 
It is worth noting that lifetimes of SMS may be shorter than the 
3\,Myr usually assigned to the most massive 
stars, and this may be relevant for studies of the upper limit of 
nascent stars that use age-related arguments \citep[e.g. ][]{figer_upper05}. 

\section{Summary and discussion}
\label{sec:concl}

In this paper we discuss the evolution of stars with initial masses in the range 
60\,\msun\, to 1001\,\msun. Our study was motivated by the results of the direct
$N$-body simulations of dense star clusters by Portegies Zwart et al.
(1999), in which a star grows by repeated collisions to well beyond
100\,\msun. In later studies, Portegies Zwart et al. (2004) proposed
that the collision runaway in star clusters could explain the presence
of a black hole of $\apgt 600$\,\msun\, in the star ULX M82~X-1 in
cluster MGG11,
which supposedly could have formed from a $\apgt 1000$\,\msun\, star.
According to these models, the mass of the VMS increases over the
stellar lifetime, starting as a homogeneous massive $\sim 100$\,\msun\,
star that grows to $\apgt 1000$\,\msun\, within the core-hydrogen burning stage of evolution.

We have assumed that our stellar evolution calculations are representative 
for stars that grow in mass via the collision runaway process. 
This is not necessarily correct, as the hydrogen reservoir of 
the collision runaway product will continuously be replenished by repeated 
collisions, whereas in our simulations we start with a high-mass homogeneous model. 
Furthermore, rotation may play a relevant role in the evolution of these objects, which 
was also ignored in our study. Having noted this, we may nonetheless expect 
these massive objects to be subject to heavy mass loss, and we may provide meaningful 
results in terms of the fate of the objects under consideration.

We have confirmed previous results on the existence of an upper mass limit for 
chemically homogeneous stars, which is reached when
the luminosity in the outermost layers of the stars approaches the
Eddington luminosity, at which point gravity is unable to balance
radiation pressure. For non-rotating solar composition stars, this limit is reached
at about 1000\,\ms.
We evolved the stars of 60\,\ms\ to 1001\,\ms\  to the end of the core helium-burning stage and calibrated 
the stellar mass-loss prescription adopted by enforcing the condition that no star should spend more 
than a few percent of its life above the HD-limit.
We found that mass-loss rates and HR-diagram positions of the earliest O-type stars, and stars classified as transitional between 
O- and WR-stars, may be consistent with our computed tracks.

Based on our mass-loss prescription and stellar evolution calculations, we argue that 
the observed massive stars in the Galaxy and the Magellanic Clouds had birth masses of up to
$\simeq200$\,\ms.  For stars in the lower
part of this mass-range, mass loss during the MS-stage leads to the
exposure of layers moderately enriched in He, and they may be observed
as late-type WN stars with hydrogen features in their spectra.  In the
upper part of this range, stars become hydrogen-deficient WR stars
already on the main-sequence.

Recently, \citet{belkus07} published a study of the evolution of stars
with masses up to 1000\,\ms.  Our study differs from that of Belkus et
al. in two important aspects: (i) they provided a simple evolutionary
recipe based on similarity theory, assuming that stars are homogeneous
throughout the entire course of their evolution due to vigorous
convection and stay in thermal equilibrium, whereas we present
detailed stellar structure models; (ii) Belkus et al. used
extrapolated mass-loss rates as predicted from radiation-driven wind
theory, whilst we employed a mass-loss prescription that is consistent with the
location of the most massive stars in the Hertzsprung-Russell diagram.
In both studies, it is found that SMS are subject to dramatic mass loss that  
probably inhibits the formation of IMBH by the runaway stellar collision scenario.

Star clusters that experience core collapse before the most massive
stars have left the main-sequence can develop a supermassive star via
collision runaway. The mass of such an object is accumulated in subsequent
collisions on a time scale of less than 3\,Myr. The mass which can be
grown in this time interval can be estimated using Eq.~(2) of
\citet{spz_etal06_imbh}\footnote{These results were obtained using standard models for runaway mass accumulation without accounting for
the possibility of vigorous mass loss as considered in the present paper.}:
\begin{equation}
  m_{\rm r} \simeq 0.01 m \left( 1 + {t_{\rm rl} \over 100\,{\rm Myr}} \right)^{-1/2},
\end{equation}
 where $m_{\rm m}$ is the mass of the object formed by runaway collision, $m$ is the 
system mass, and $t_{\rm rl}$ is the system relaxation time. The average mass increase per collision is about $\sim 20$\,\msun\
\citep{pmmh99} (for a more complete discussion, however,
see \citet{pm_imbh_clust02}). The supermassive star that
accumulates $\sim 1000$\msun\, has experienced some 45
collisions between the moment of gravothermal collapse of the cluster
and the moment that the supermassive star dies. The
mean time between collisions for this model is $\aplt 6.0 \times
10^4$\,years, resulting in a mass accretion rate of $\apgt 3\times
10^{-4}$\,\msun/yr. 
This rate of mass accretion is lower than the rate of mass loss that 
we get for the most massive stars ($\sim$ $3.8\times 10^{-3}$\,\msun/yr), but since 
these numbers of mass-accretion and mass-loss rate are quite uncertain, it is not inconceivable that, in spite of the strong 
stellar winds, the objects may nonetheless experience a net gain in mass during their lifetimes.
This conclusion was drawn by \citet{suzuki07} who recently studied stellar evolution with mass loss 
of collisionally merged massive stars. They concluded that stellar winds would not inhibit 
the formation of SMS. We note, however, that the Suzuki et al. (2007) calculations
were limited to masses up to $\sim$ 100 \ms\ and their calculations did not consider
the more important effect of mass loss for the final mass growth process, where the mass 
is supposed to increase beyond this value of 100 \ms\ by an order of magnitude.

In this study, we have found that a super-massive star is likely to shed most of
its mass well before it experiences a supernova explosion.  In several cases the
supernova progenitor still had $\apgt 100$\,\msun, and such objects could
possibly collapse to black holes of intermediate mass but with $M_{bh} <<
1000$\,\ms. Our calculations suggest the possibility of the formation of objects
that experience pair-instability supernovae, which would be interesting  
phenomena to observe. The recently discovered, very luminous supernova 2006gy may have 
been a PISN \citep{smith_07,ofek_07,langer_07}, although 
this is still under debate.

Obviously there is the possible caveat that the quantitative difference between
starting as a homogeneous high-mass star (as we considered in our paper), or
growing to one  over the main-sequence lifetime by repeated collisions may be
significant, but at this stage we cannot provide further insights about the
consequences of these potential differences.  We therefore draw our conclusions
on the calculations at hand, and we argue that  the majority of supermassive
stars probably end up as black holes of $M \lesssim 70$\,\ms, with the possible 
exception of some of them exploding as pair-instability supernovae.   With the
approximate nature of our applied mass-loss algorithm in mind, and also taking
into account the (also approximate) results of \citet{belkus07}, we infer that
the accuracy on the limit for black hole masses is $\sim$ $\pm 30$\,\ms. We
therefore conclude that most supermassive stars end their lives as $70\pm 30
$\,\msun\, black holes. 

\section*{Acknowledgments}

We thank the referee, A. Maeder, for constructive comments that helped improve the paper.
We also thank L. Kaper, G.-J. Savonije, N. Langer, Yu. Fadeyev, N. Chugai, D. Vanbeveren
and O. Pols for productive discussions.  LRY acknowledges
P.P. Eggleton for providing his evolutionary code and advice on its
usage. Z. Han is acknowledged for sharing his knowledge of this
code. We thank F. Martins  and G. Rauw
 for providing us the data 
 in
advance of publication. LRY is supported by NWO (via grants
\#635.000.303 and \#643.200.503), NOVA, the Russian Academy of
Sciences Basic Research Program ``Origin and Evolution of Stars and
Galaxies'' and the Russian Foundation for Basic Research grant \#07-02-00454.  
LRY acknowledges warm hospitality and support from
the Astronomical Institute ``Anton Pannekoek'', where most part of this
study was carried out.

\section*{APPENDIX:  Notes on individual stars} 

We first discuss the enigmatic LBV $\eta$~Car. 
We adopt a luminosity of $5 \times 10^6$\,\ls, which 
corresponds to the observed IR-flux and assumed distance of 2.3 Kpc  \citep{hillier01}. 
The positions of the two labels for the star in the HR-diagram reflect the ambiguity 
in the determination of the stellar radius, caused by the presence of an optically thick wind that dominates 
the spectrum and prevents the determination of $R_\star$. 
The maximum and minimum \te\ correspond to a Rosseland optical depth of $\tau$ = 155 and 0.67 respectively in the atmosphere.   
$\eta$~Car may be a $2020 \pm 5$ day binary \citep{dam_etacar97,damineli_etacarbin00}. 
Assuming that the luminosity is Eddington, Hillier et al. (2001) estimate the minimum total mass of the system to be 
$150[L/(5 \times 10^6)]$\,\ms), 
but some contribution of the secondary star is probably not very significant, although the secondary is possibly a 
late-O/WR star \citep{iping_ecar_bin05}. 
Our computations show that the luminosity of a several hundred solar mass star may still be far 
from the Eddington limit. The estimated mass-loss rate is $\md \approx 10^{-3}$\,\myr. 
Bearing in mind that $M_0=(200 - 300)$\,\ms\ stars in their core-hydrogen burning stage evolve at almost constant luminosity, 
$L = 5 \times 10^6$\,\ls\ in our computations corresponds to a ZAMS mass of about 250\,\ms. 
\citet{hillier01} derive for the $\eta$~Car atmosphere a H/He fraction of $\approx$5 by number, i.e. X$\approx$0.56, Y$\approx$0.44. 
Thus, $L$, X, Y, \md, and the ``high'' \te\ solution suggest that $\eta$~Car 
is a $\sim 250$\,\ms\ star somewhere in the middle of the core-hydrogen burning stage. 
The nature of its instability has yet to be found.

The position of the candidate LBV Pistol star is drawn according to the ``low-luminosity'' solution of 
\citet{figer_pistol98}, since \citet{naj_pistol99} found that its metal content equals $Z\simeq 3 Z_\odot$, 
corresponding to its location in the star-forming region of the Galactic center, and that this favours the lower 
luminosity solution for its stellar parameters. Note however that, if it turns out that the Pistol star is an LBV, its $T_{\textrm eff}$ will be variable. 
Of all our tracks, the one for $M_0=200$\,\ms\ seems to provide the best fit of the position of the Pistol star.
\citet{figer_pistol98} find Y=0.3-0.4 at the surface of the Pistol star, as well as strong N enhancement. 
Going redward, at $\log T_{\rm eff}=4.15$, our $M_{0}=200$\,\ms\ evolutionary sequence has a mass $M=92$\,\ms\ and  Y=0.86. 
Going blueward, it has a mass of M=86.5\,\ms\ and Y=0.91. 
We therefore cannot match the chemical composition of the Pistol star. 
Other tracks passing through the same position also have too high a helium fraction $Y$. 
For instance, the track for $M_{0}=120$\,\ms, which runs slightly lower than the error box of the Pistol star, 
has Y$>0.7$. Note that the determination of the He-abundance is challenging.
However, if the high metallicity of the Pistol star is real, it may be an almost unevolved star: \citet[][ see their Fig.~6]{ishiietal99} find 
that the Pistol star lies very close to the ZAMS for stars with $X$=0.7, $Z$=0.1; its mass is then slightly higher 
than 200\,\ms\footnote{We must note however that the models of \citet{ishiietal99} for mass higher than $\simeq 200$\,\ms\ have more extended envelopes than our models.}. 

There is another LBV star in the Quintuplet cluster, similar to the Pistol star: FMM362 \citep{figer_massive_center99,geballe_fmm362}. 
The latter authors report $T_{\rm eff}= (10000 - 13000)$\,K for this object, with $L \geq 10^6$\,\ls.

For the stars in the HeI cluster in the Galactic centre, we show \md\ estimates from
\citet{martins_gc07}.  
   The positions of the
HeI stars in the HR-diagram are consistent with their interpretation as being
evolved blue supergiants,  close to the WR stage.  We note that the evolutionary
stage of the HeI star has to be very short.

We note that the WN stars -- termed AdK -- may be in their core-hydrogen burning stage, as they are positioned along 
portions of the tracks where Y is only moderately enhanced.

The cool hypergiant IRC~+10~420 is the only object that is believed to be observed in the phase of rapid transition from 
the red supergiant stage to the WR phase \citep[see e.g.][ and references therein]{humphreys_irc97,block_hyp99}.
Its $T_{\rm eff}$ appears to have increased by 1000 -- 2000K within the last 20 years. 
We note that 
\citet{smith_vink_koter04} computed radiation-driven wind models for objects with a low stellar mass and discussed 
the possibility that these cool hypergiants like IRC~+10~420 may in fact be ``LBVs in disguise'', where 
a large mass-loss rate induced by the high Eddington factor produces a pseudo-photosphere.
Although the object is located in the HR-diagram at the lowest luminosity level, based on our calculations we expect the 
existence of similar more luminous stars, which have to be extremely rare due to their short lifetimes. 
$\rho$ Cas may be an example 
of star that is in an unstable post-RSG stage of evolution, when it experiences outbursts ejecting mass at a rate 
of several 0.01\,\myr\ \citep{lobel_rhocas03}. 

We show the position of one of the brighter LBVs, AG Car, after \citet{hum_dav94} and \citet{lamers96a}. 
Two data labels illustrate the range of excursions over the HR-diagram experienced by the LBV. 

Especially remarkable is WR20a, a $P_{\rm orb} \approx 3.68$ day massive
binary \citep{rauw+04,rauw+05,bonanos_wr20a} with both components of a WN6ha spectral type (see Fig.~\ref{fig:hrd_obs_col025}). 
The  component mass estimates are $83\pm5.0$\,\ms\ and 
$82\pm5.0$\,\ms\ \citep{bonanos_wr20a}. 
This makes these components of WR20a the most massive stars weighed in binaries. 
According to \citet{rauw+05}, fundamental parameters of each of the components are: $\te = 43000 \pm 2000$\,K, 
$L_{\rm bol}/\ls \simeq (1.15 \pm 0.15) \times 10^6$, 
$\dot{M} = 8.5 \times 10^{-6}$\,\myr  (assuming a clumped wind with a volume filling factor $f = 0.1$). 
Nitrogen is enhanced in the atmospheres, whilst carbon is depleted. 
Spectral classification implies only a weak helium enrichment of the atmosphere. The origin and evolution of this system deserves urgent study. 
\citet{rauw+05} propose that the position of the WR20a components in the HR-diagram ``suggests that they are 
core hydrogen burning stars in a pre-LBV stage and their current atmospheric chemical composition probably 
results from rotational mixing that might be enhanced in a close binary compared to a single star of the same age''. 
However, we should note that the evolution with complete mixing of a high-rotational velocity star in a 
synchronised binary with mass loss may result in evolutionary tracks that evolve to the left of the ZAMS, in 
apparent conflict with the position of WR20a. 
As Figs.~\ref{fig:mmdot} and \ref{fig:hrd_obs_col025} show, \te, L, and $\mathrm{\dot{M}}$ of WR20a are actually 
quite consistent with a star with an initial mass of about 100\,\ms\ in a rather early stage of core-hydrogen burning, 
when the layers enriched in He are beginning to be exposed.    

\begin{figure*}
\includegraphics[angle=-90,scale=0.5]{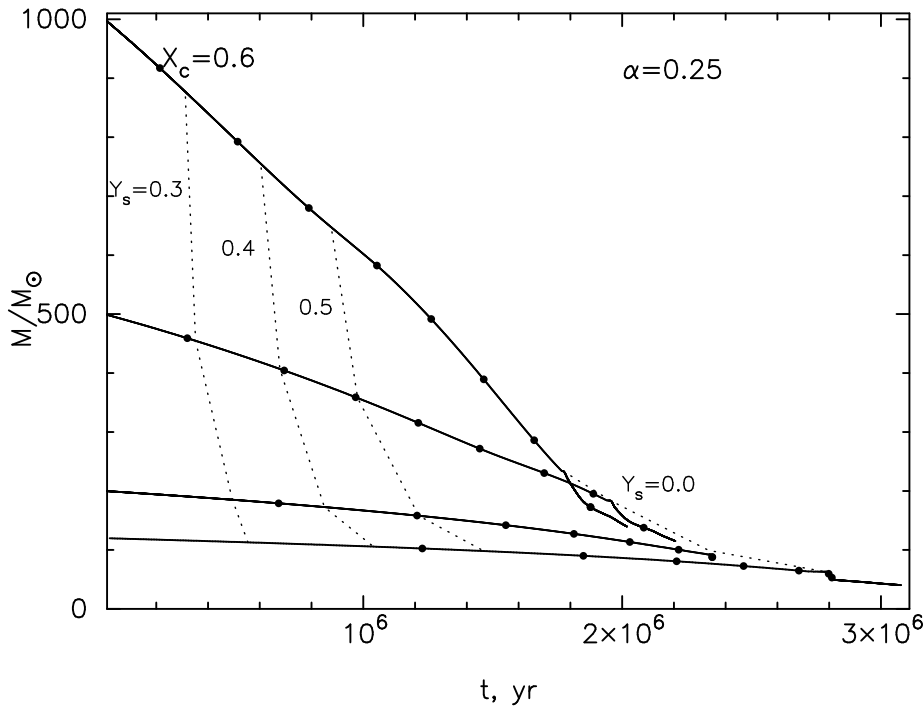}
\caption[]{Stellar mass vs. evolutionary lifetime for $\alpha$=0.25-sequences of stellar models. 
The dotted lines connect the loci of models with surface helium abundance 0.3, 0.4, 0.5, and 0.0.
The dots at the curves indicate the time and mass when the hydrogen abundance in the convective cores ${\rm X}_{\rm c}$ becomes 0.6(0.1)0.}
\label{fig:tmhy025}
\end{figure*}

\begin{figure*}
\includegraphics[angle=-90,scale=0.5]{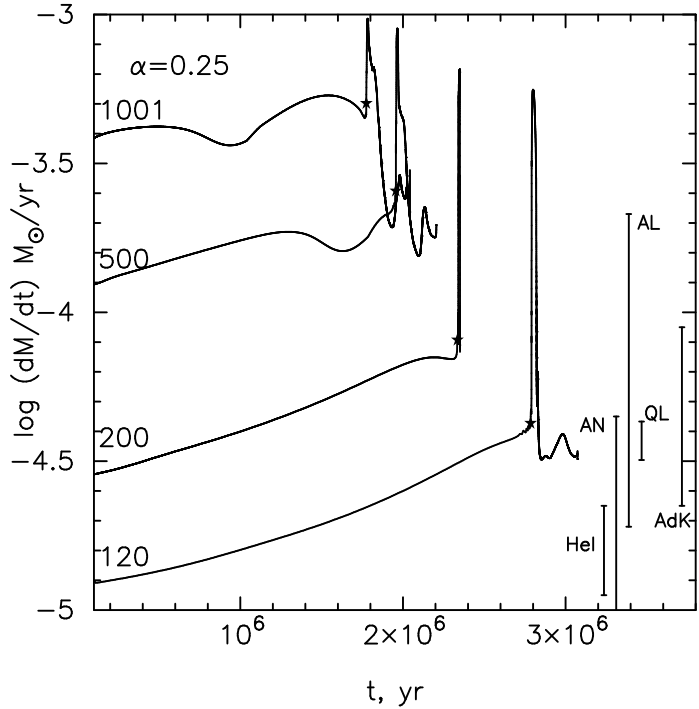}
\caption[]{Mass-loss rate {\textit vs.} evolutionary lifetime for models with  $\alpha$=0.25. 
The left border of the plots corresponds to $t=10^5$\,yr, since mass-loss rates before this time are virtually the same as for the 
left extremes of the curves. The star symbols at the curves indicate the central hydrogen exhaustion time. 
Vertical bars show the ranges of the estimates of the mass-loss rate, obtained by means of quantitative spectroscopy for the 
Arches cluster \citep[][ AN]{najarro_arches_met04}, HeI emission stars in the Galactic center
\citep{martins_gc07},
star-forming-regions in the Galaxy and the Magellanic clouds (AdK, A.de Koter, unpublished), and 
estimates for the Arches and Quintuplet stars obtained by radio-observations  
\citep[AL and QL, respectively, ][]{lang05}.   
For the Quintuplet, only \md\ for stars identified with stellar winds are shown. 
In the Arches cluster, \citet{lang05} identify stellar winds for all observed sources. 
}
\label{fig:tdmdt025}
\end{figure*}

\begin{figure*}
\includegraphics[angle=-90,scale=0.5]{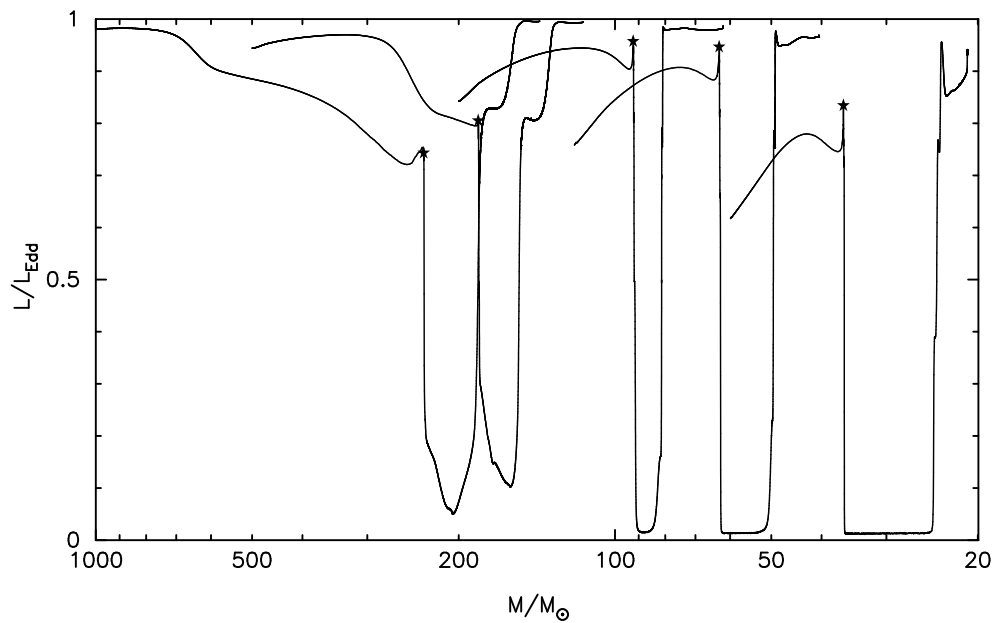}
\caption[]{The variation of $L/L_{\rm Edd}$ in the outer meshpoint of the models along evolutionary sequences 
for stars with $M_0$=1001, 500, 200, 120, and 60\,\ms, for $\alpha$=0.25. The star symbols label models with 
hydrogen abundance in the convective core $X_c=0$.
}
\label{fig:lledd_025}
\end{figure*}

\begin{figure*}
\includegraphics[angle=-90,scale=0.5]{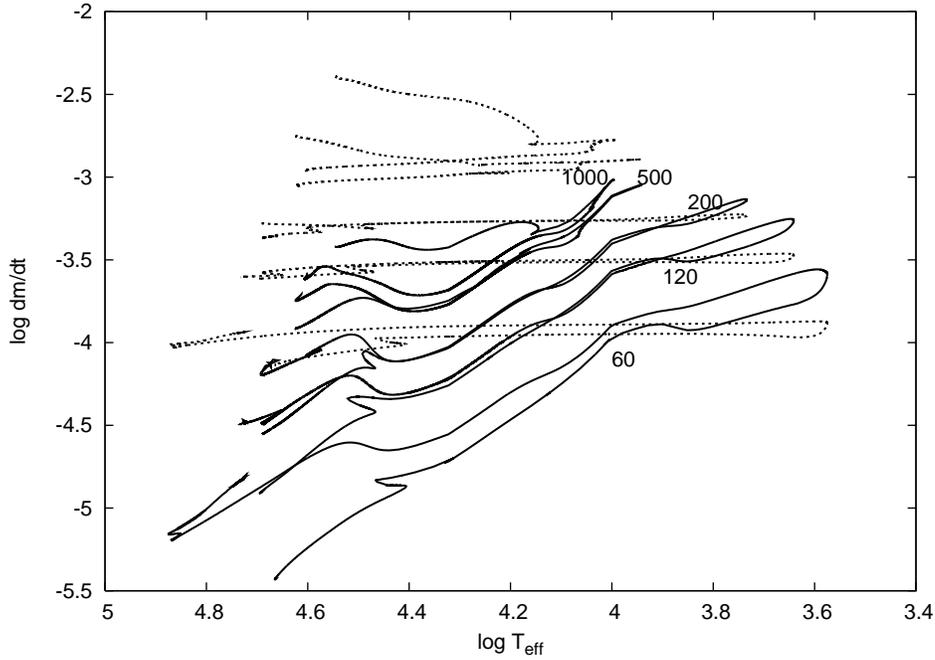}
\caption[]{The variation of the mass-loss rate versus $T_{\rm eff}$ for the $\alpha$=0.25 sequence 
of stars (solid curves).
For comparison, the 
dotted curves
represent estimates of the maximum \md\ along 
the same tracks according to Eq.~\ref{eq:dmdt_max}.}
\label{fig:te_dmdt_025}
\end{figure*}

\begin{figure*}
\includegraphics[angle=-90,scale=0.5]{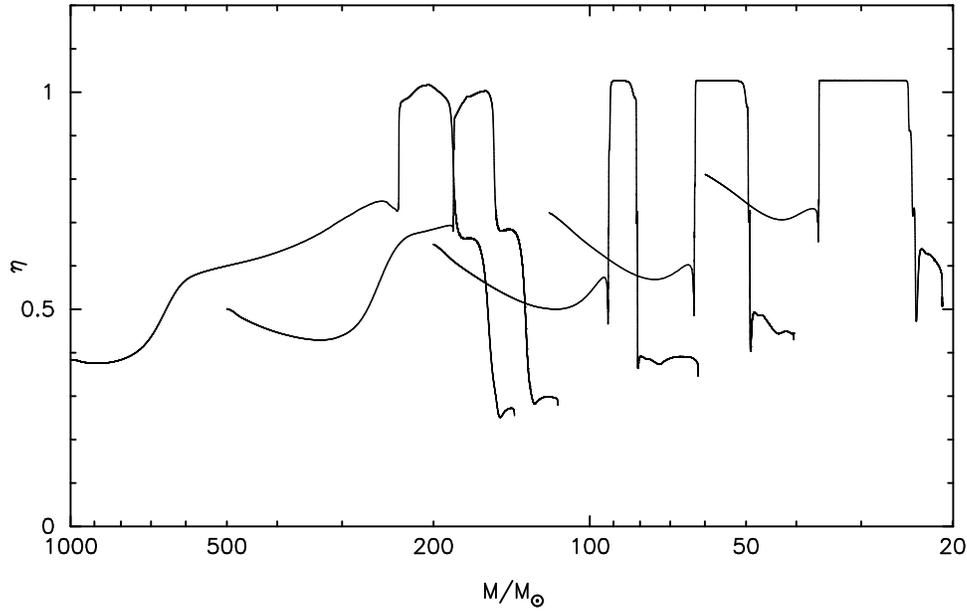}
\caption[]{The variation of the ``wind performance number''  along evolutionary 
tracks for $\alpha$=0.25.}
\label{fig:perfn}
\end{figure*}

\begin{figure*}
\includegraphics[angle=-90,scale=0.7]{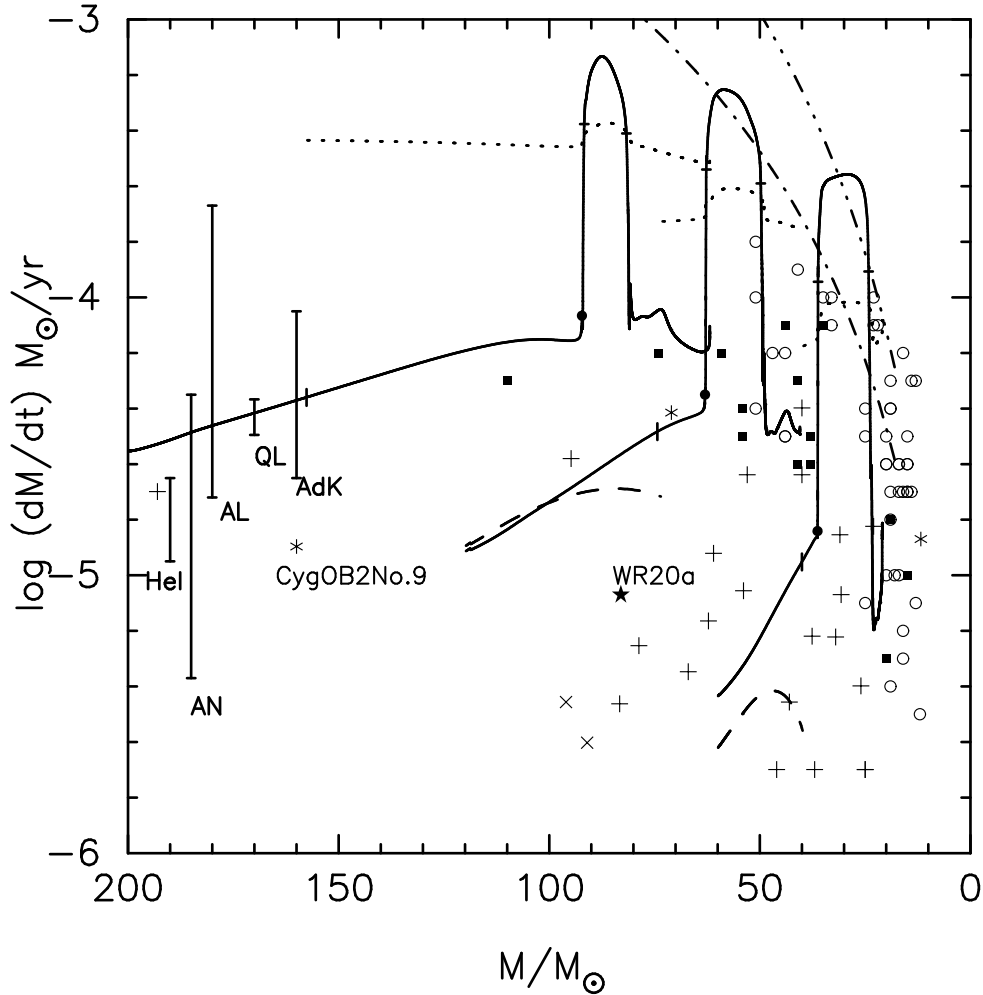}
\caption[]{The relation between stellar mass and mass-loss rate for the
$\alpha$=0.25 case.
Ticks along the tracks (in the order of decreasing mass) indicate models in
which the
surface He-abundance becomes $Y_s=0.4$, the last model with $T_{\rm eff}
\geq 10000$\,K and $Y_s \geq 0.4$, and the
first model in which  $T_{\rm eff} \geq 10000$\,K again.
The dots at the tracks mark the TAMS.
The long-dashed line shows the mass-loss rate computed using the
\citet{vkl01} mass-loss formula.
Dotted, dot-dashed, and dot-dot-dot-dashed  lines show mass-loss rates for
hydrogen-deficient WR  stars from
\citet{donder_vanbev_wr03}, \citet{lan89b}, and \citet{nelemans01}, respectively.  Black squares
represent estimates of stellar parameters of observed hydrogen-rich WR stars
based on model atmospheres that 
account for iron-line blanketing and clumping \citep{hamann_wn06}: open
circles are estimates by the same authors for hydrogen-poor
stars ($X_H<0.2$)
Asterisks are spectroscopic mass and \md\ estimates for Galactic O-stars,
plusses are estimates
for LMC O-type stars, crosses are estimates for SMC stars (see Table~\ref{tab:obs} for references).
The star symbol marks the position of the almost identical components of WR20a, 
currently the most massive star weighed
in a binary system. Note its low \md.  Vertical bars have the same
meaning as in Fig.~\ref{fig:tdmdt025}.
}
\label{fig:mmdot}
\end{figure*}

\begin{sidewaysfigure*}
\centering
\includegraphics[angle=-90,scale=0.60]{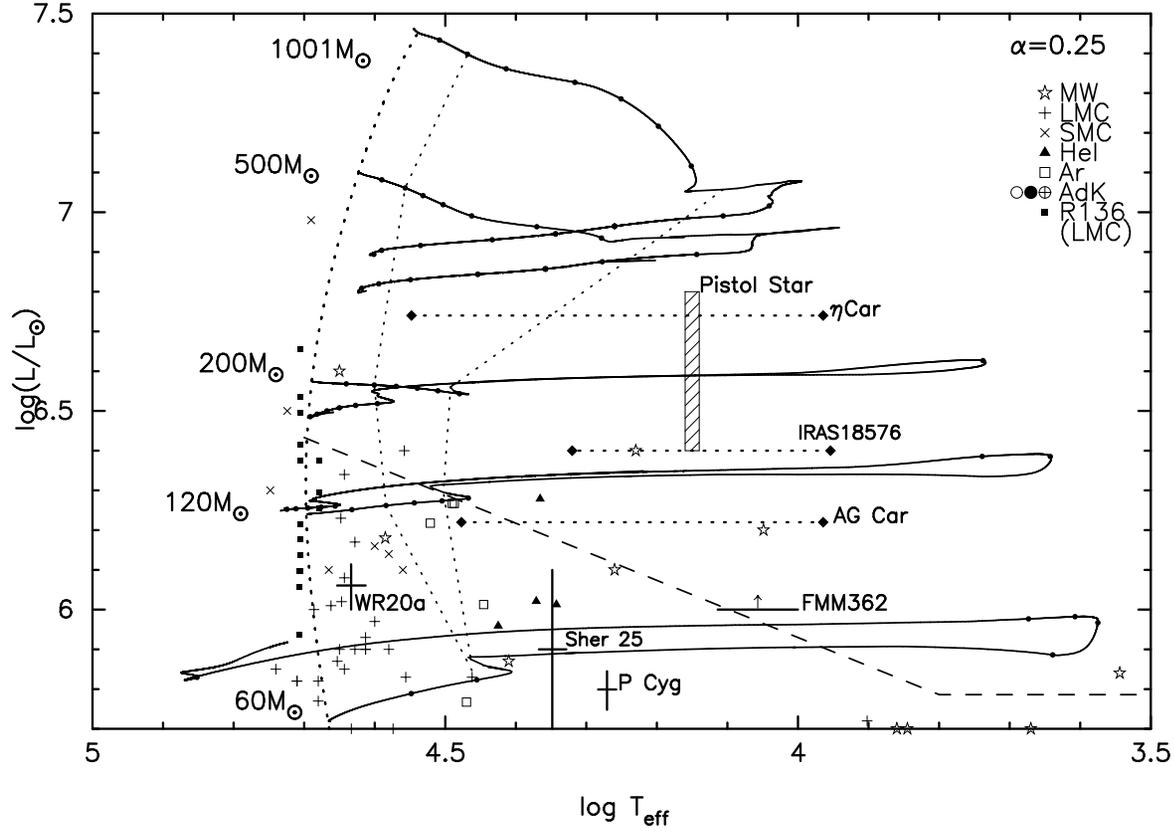}
\caption[]{Evolutionary tracks of stars in the Hertzsprung-Russel diagram (solid lines, $\alpha$=0.25 case). Dots along the tracks show positions of models with 
surface helium abundance $Y_s$=0.3(0.1)0.9. 
Three  dotted lines stretching from top to bottom connect ZAMS stars, 
 the loci of models with surface He-abundance $Y_c=0.4$  (models with higher $Y_c$ are conventionally considered as WR-stars), and TAMS stars. 
Broken long-dashed line shows the Humphreys-Davidson limit (linearly extrapolated at high L and $T_{eff}$). 
The hatched rectangle shows the position of the Pistol star (see the text for comments).
Also plotted are the positions of the most luminous stars in the Galaxy, the LMC, and the SMC (Table~\ref{tab:obs}), the HeI-emission line stars in the Galactic 
center \citep{najarro_gc97}, the Arches cluster stars \citep{najarro_arches_met04}, luminous WNLh stars in the Galaxy and the Magellanic clouds 
(AdK, A. de Koter, unpublished) and the most luminous stars in the R136 cluster, with spectral subtypes O3 If*, WN, O4If, and O3 III(f*) \citep[][ Table3]{massey_hunter98}. 
For AdK-stars, open circles are for the Galaxy, filled circles are for the SMC, crossed circles are for the LMC. 
See Appendix for the notes on particular stars.  }
\label{fig:hrd_obs_col025}
\end{sidewaysfigure*}

In addition to the Galactic stars, we  plot some Magellanic Clouds stars in Fig.~\ref{fig:hrd_obs_col025}.
Most remarkable are the R136 cluster members of  O3(If*) and
O3If*/WN6A subtypes that are considered to be transitional between O and WN type stars \citep{massey_hunter98}\footnote{The degeneracy 
with respect to the effective temperature is caused by the absence of individual determinations of $T_{\rm eff}$. 
The latter were assigned according to spectral type  after \citet{vacca96}.}.
The stars are located along the ZAMS or slightly to the left of it. 
The R136 cluster belongs to the LMC and has a several times lower $Z$ than Galactic stars. 
Figure 6 of \citet{ishiietal99} shows that in the (100--200)\,\ms\ range, ZAMS stars with 
$Z$=0.004 are $\sim 0.1$\,dex in $\log T_{\rm eff}$ hotter than $Z$=0.02 stars, but they have the same luminosity.  
As our Fig. \ref{fig:hrd_obs_col025} shows, (100-200)\,\ms\ stars may easily reveal over-abundances of helium, whilst they evolve 
over $\Delta \log T{\rm eff} \sim 0.1$ in the HR-diagram. 
The surface helium abundance increases to $Y=0.4$ in $\sim 0.5$\,Myr.  
Therefore, R136 cluster members may be very massive stars that are still in their core H-burning phase, with a strong 
stellar wind that causes the spectrum to mimic that of evolved WR stars, as was originally conjectured for one of 
these stars (R136-006$\equiv$R136a3) by \citet{koeteretal97}. 
Another possibility is that the surface layers enriched in helium are already exposed (this may 
happen in $\sim 1$\,Myr, see Fig. \ref{fig:tmhy025}). This is also consistent with the star formation history 
of R136 \citep[see the discussion in ][]{massey_hunter98}. 
We also note that the effective-temperature scale by \citet{martins_etal06} assigns 
O3I-type stars  $\log T_{eff} \approx 4.6$ (instead of.4.7 in Vacca et al. 1996) and smaller bolometric 
corrections (implying a reduction of the luminosity by $\approx$ 0.25 dex).
This would move these stars to the right of the ZAMS, in the proper direction.
The position of R136 members 
in the HR-diagram is roughly consistent with the estimate of the upper limit of masses of stars 
in this complex of (140--160)\,\ms\ obtained by \citet{koen_r136} by power-law distribution fitting to these stars. 

\begin{sidewaysfigure*}
\centering
\includegraphics[angle=-90,scale=0.63]{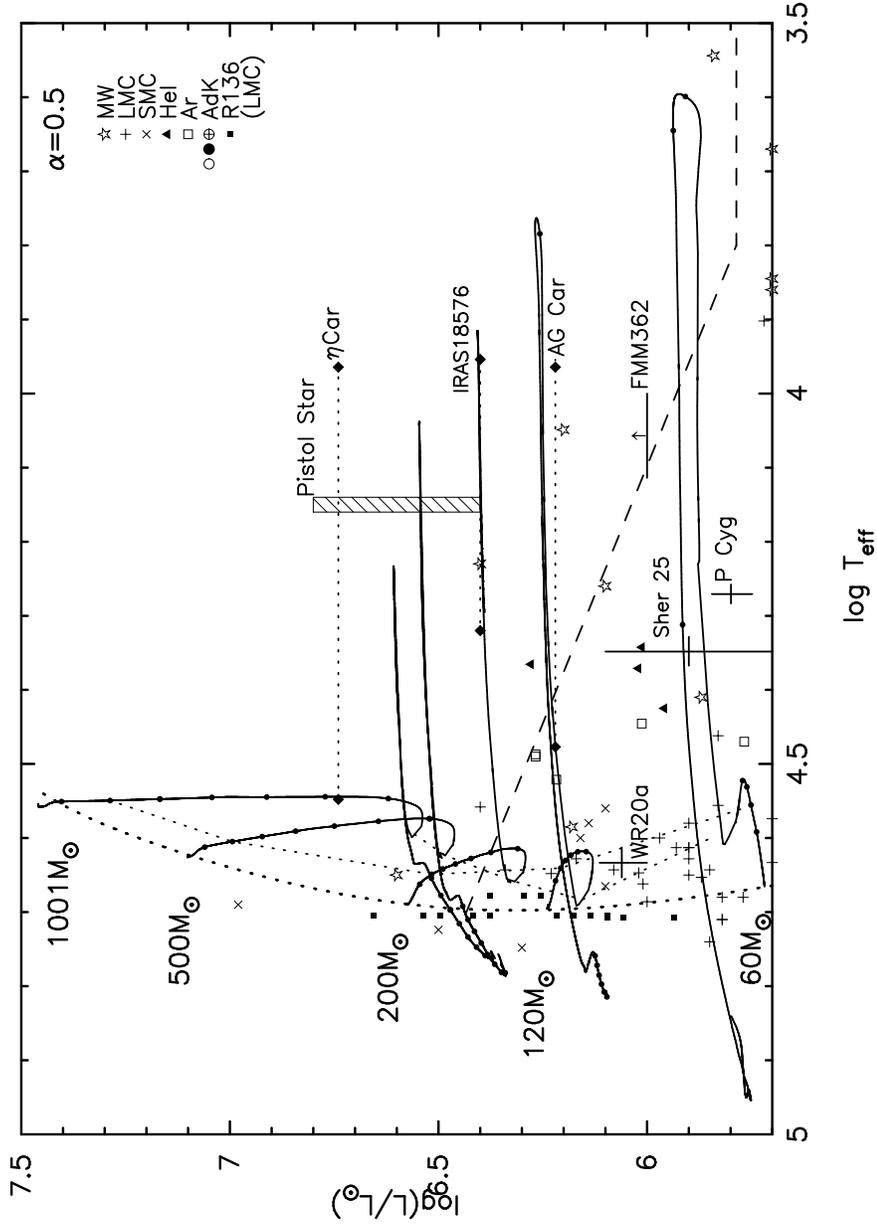}
\caption[]{Same as Fig.~\ref{fig:hrd_obs_col025} but for $\alpha$=0.5.}
\label{fig:hrd_obs_col05}
\end{sidewaysfigure*}

\begin{figure*}[t!] 
\vskip -3cm
\hskip -2.5cm
 \includegraphics[]{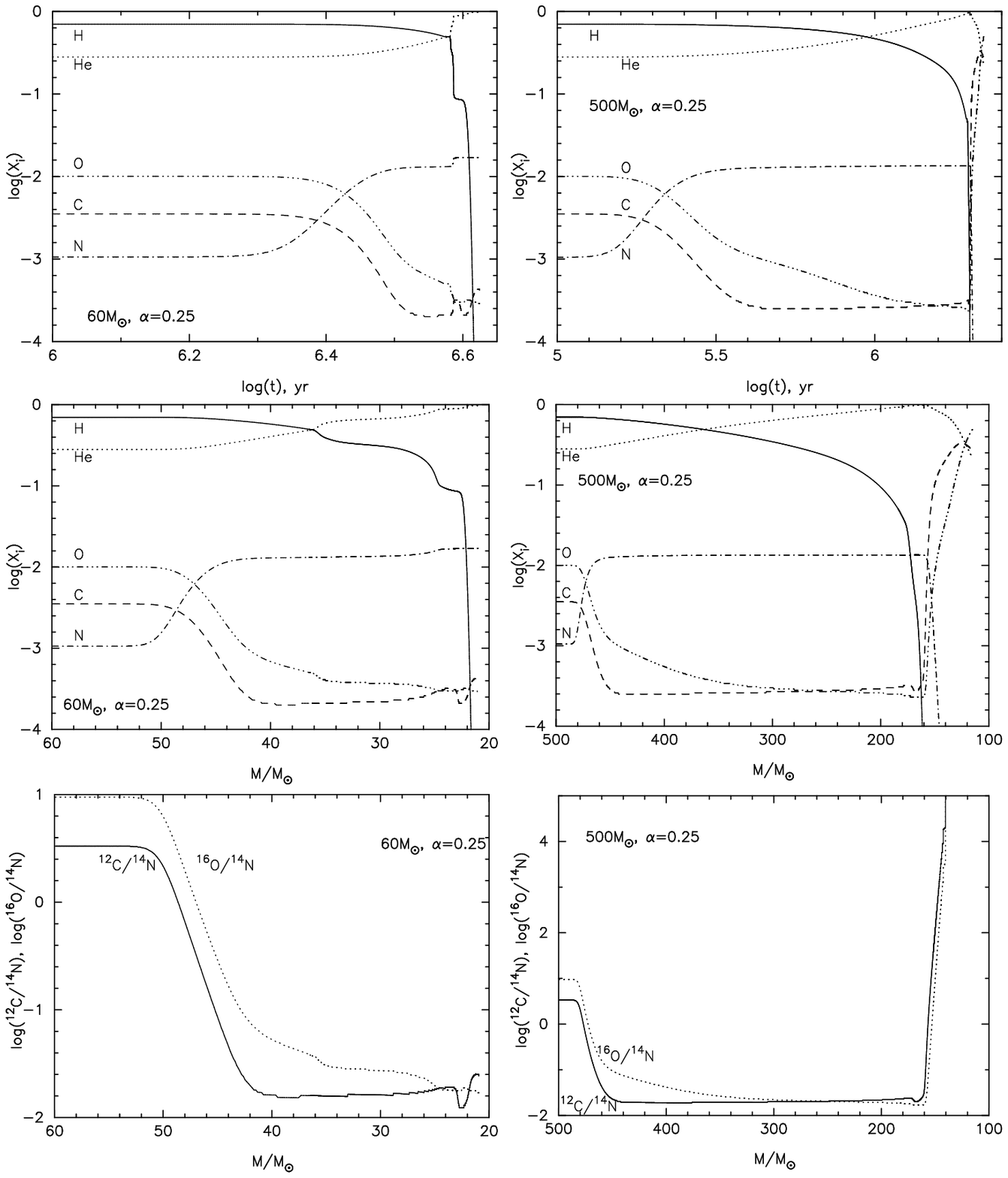}
 \vskip -2cm
  \caption{Variations of the surface abundances vs. time (upper panels) and mass (middle
  panels), and the variations in the surface isotope ratios vs. mass (lower panel) for models of $M_0$=60 (VMS) and
  500\,\msun\ (SMS) stars.}
	\label{fig:chem}
\end{figure*}

\begin{figure*}
\centering
\includegraphics[angle=-90,scale=0.7]{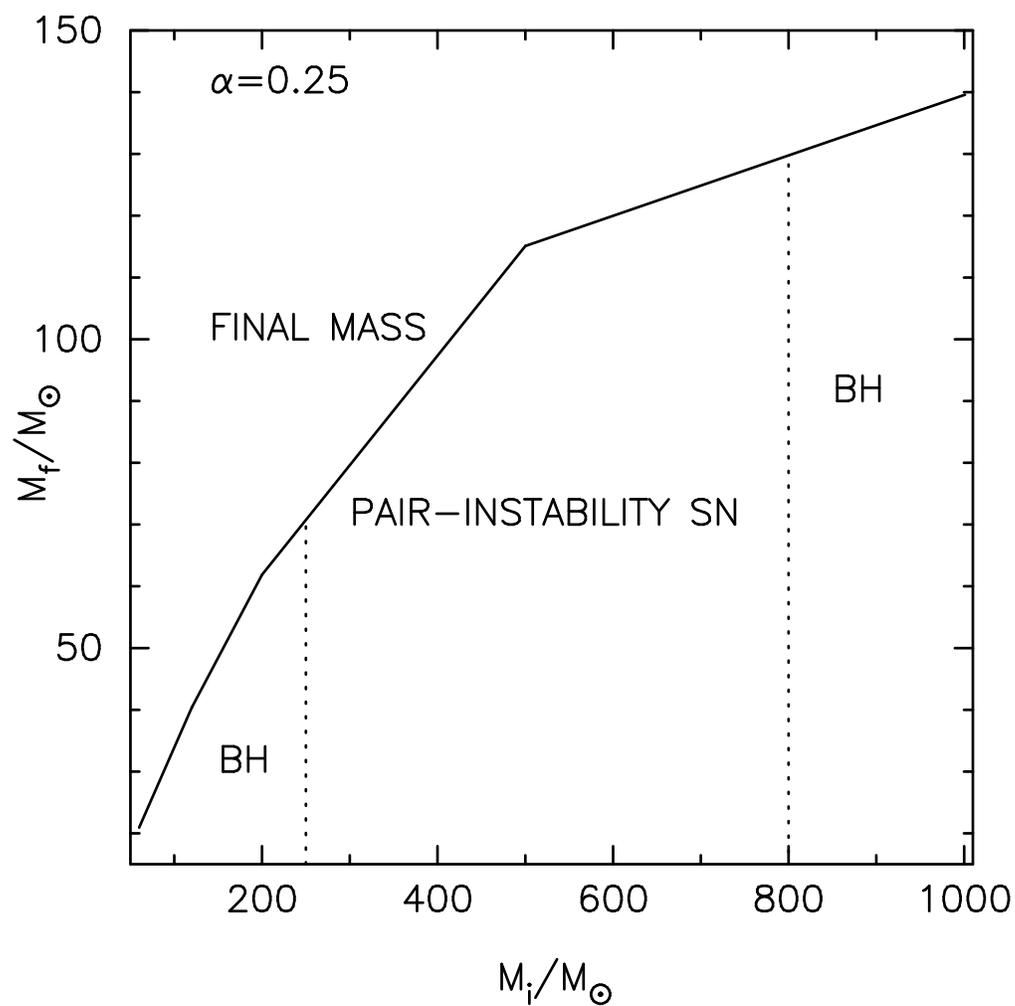}
\caption[]{Initial-final mass relation for SMS and inferred outcomes of evolution: black hole formation 
for progenitors with mass $M_i\aplt 250\,$\ms\ or $M_i\apgt 800\,$\ms, and pair-instability supernovae in the intermediate range of $M_i$.}
\label{fig:mimf}
\end{figure*}

\begin{figure*}
\centering
\includegraphics[angle=-90,scale=0.7]{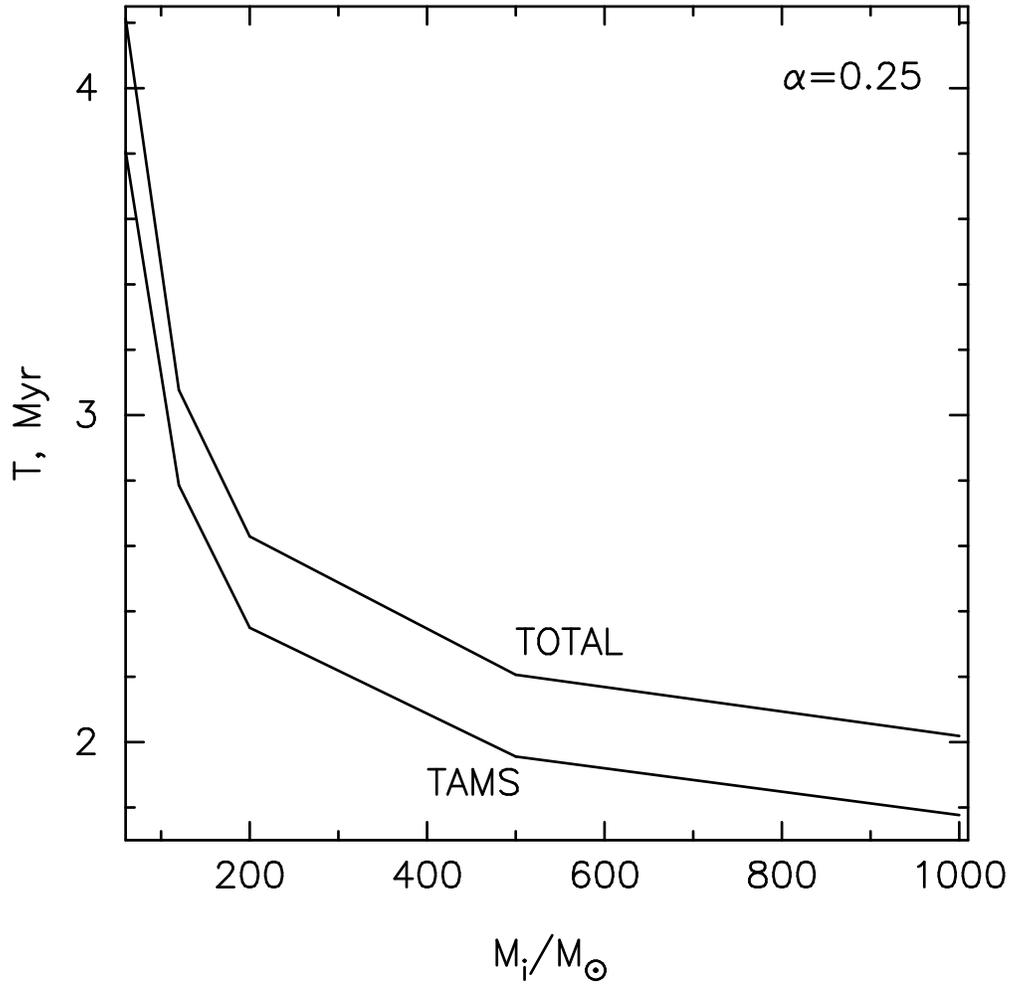}
\caption{Total lifetimes of supermassive stars and their lifetimes in the core-hydrogen burning stage.}
\label{fig:life}
\end{figure*}

Finally, we also plot P Cygni \citep{lamers96b} in Figs.~\ref{fig:hrd_obs_col025} and \ref{fig:hrd_obs_col05}, 
and show the positions of several other high-luminosity Galactic stars for which our calculations 
may be relevant (see Table~\ref{tab:obs} to identify them if they are not annotated in the Figures).

\pagebreak
\bibliographystyle{aa}

\bibliography{used_jorick}

\newpage
\begin{table}[!ht]
\begin{onecolumn}
\caption[]{Summary of calculated evolutionary tracks. Successive columns give the initial mass $M_0$, the value of the $\alpha$ parameter, the age at the 
terminal age main-sequence $t_{\rm TAMS}$, the mass at the TAMS $M_{\rm TAMS}$, the age of the last computed model $t_{\rm f}$, the mass of the last computed 
model $M_{\rm f}$, the central abundances of He, C, and O in the last model $Y_{\rm cf}$, $X_{\rm C,cf}$, $X_{\rm O,cf}$, and the abundances of H and 
He
at the surface of the last computed model  $X_{\rm sf}$, $Y_{\rm sf}$.
Note that helium burning results in the formation of oxygen-neon cores with traces of carbon. The ``V ''  in the second column stands for 
the 60\,\ms\ track 
computed using the mass-loss prescription of \citet{vkl00}.   
}
\centering
\begin{tabular}{lllllllllll}
\hline\hline
$M_0$ & $\alpha$ & $t_{\rm TAMS}, 10^6$ yr & $M_{\rm TAMS}$ & $t_{\rm f}, 10^6$ yr & $M_{\rm f}$  & $Y_{\rm cf}$       & $X_{\rm C,cf}$ & $X_{\rm O,cf}$  & $X_{\rm sf}$ & $Y_{\rm sf}$  \\ 
1001  & 1.0      & 2.492		   & 29.66	    & 2.594		   & 25.24	  & 0.648	       & 0.287        & 0.039	      & 0.0	     & 0.976	        		 \\ 
1001  & 0.9      & 2.360		   & 34.82	    & 2.713		   & 20.25	  & 0.018	       & 0.141        & 0.811	      & 0.0	     & 0.204	        	\\ 
1001  & 0.8      & 2.244		   & 41.46	    & 2.541		   & 25.11	  & 0.082	       & 0.262        & 0.630	      & 0.0	     & 0.286	           \\ 
1001  & 0.5      & 1.935		   & 86.81	    & 2.218		   & 49.17	  & 0.0035	       & 0.061        & 0.856         & 0.0	     & 0.174	        	\\ 
1001  & 0.3      & 1.792		   & 219.9	    & 1.793		   & 218.4	  & 0.981	       & 0.00028      & 0.00022       & 0.025	     & 0.958	           \\ 
1001  & 0.25     & 1.777		   & 229.0	    & 2.019		   & 139.6	  & $9.4\cdot10^{-4}$  & 0.0	      & 0.819	      & 0.035	     & 0.200	       \\ 
844   & 1.0      & 2.513		   & 29.56	    & 2.899		   & 13.61	  & 0.0 	       & 0.115        & 0.844	      & 0.0	     & 0.030	        	 \\ 
598   & 1.0      & 2.572		   & 29.20	    & 2.591		   & 27.91	  & 0.953	       & 0.021        & 0.0004        & 0.038	     & 0.943	          \\ 
500   & 1.0      & 2.605		   & 28.96	    & 2.623		   & 26.96	  & 0.948	       & 0.024        & 0.00045       & 0.033	     & 0.948	           \\ 
500   & 0.9      & 2.483		   & 33.91	    & 2.837		   & 20.09	  & 0.019	       & 0.145        & 0.806	      & 0.0	     & 0.213	       \\ 
500   & 0.8      & 2.372		   & 40.02	    & 2.696		   & 23.86	  & 0.046	       & 0.195        & 0.731	      & 0.040	     & 0.239	           \\ 
500   & 0.5      & 2.104		   & 75.76   	    & 2.393		   & 46.95	  & 0.0	               & 0.062        & 0.852         & 0.0	     & 0.195	          \\ 
500   & 0.3      & 1.981		   & 163.7	    & 1.982		   & 163.3	  & 0.981	       & 0.00045      & 0.00023       & 0.057	     & 0.923	          \\
500   & 0.25     & 1.956		   & 183.4	    & 2.206		   & 115.1	  & $0.41\cdot10^{-3}$ & 0.0	      & 0.827	      & 0.039	     & 0.215	          \\  
200   & 1.0      & 2.951		   & 26.76	    & 2.997		   & 25.74	  & 0.960	       & 0.014        & 0.00032       & 0.051	     & 0.930	         \\
200   & 0.9      & 2.837		   & 30.75	    & 3.208		   & 18.98	  & 0.010	       & 0.128        & 0.828	      & 0.0	     & 0.228	         \\
200   & 0.8      & 2.740		   & 35.51	    & 3.083		   & 22.31	  & 0.027	       & 0.157        & 0.786	      & 0.0	     & 0.238	         \\
200   & 0.5      & 2.511		   & 56.71	    & 2.525		   & 54.61	  & 0.981	       & 0.0004       & 0.00024       & 0.047	     & 0.934	         \\
200   & 0.3      & 2.379		   & 81.92	    & 2.397		   & 76.12	  & 0.980	       & 0.0004       & 0.00024       & 0.073	     & 0.907	         \\ 
200   & 0.25     & 2.350		   & 92.28	    & 2.629		   & 61.88	  & 0.0 	       & 0.00556      & 0.846         & 0.0	     & 0.304	         \\
120   & 1.0      & 3.345		   & 24.15	    & 3.360		   & 23.67	  & 0.974	       & 0.0012       & 0.00027	      & 0.074	     & 0.398	         \\ 
120   & 0.5      & 2.906		   & 44.36	    & 2.915		   & 44.19	  & 0.975	       & 0.00070      & 0.00025       & 0.094	     & 0.887	          \\
120   & 0.4      & 2.836		   & 50.74	    & 2.844		   & 50.52	  & 0.980	       & 0.00037      & 0.00025       & 0.133	     & 0.848	         \\ 
120   & 0.35     & 2.813		   & 54.41	    & 3.108		   & 37.86	  & 0.981	       & 0.00036      & 0.00025       & 0.163	     & 0.818	         \\
120   & 0.3      & 2.771		   & 58.79	    & 3.080		   & 39.34	  & 0.847	       & 0.073        & 0.847	      & 0.0	     & 0.320	         \\
120   & 0.25     & 2.786		   & 63.02	    & 3.077		   & 40.38	  & 0.043	       & 0.146        & 0.781	      & 0.0	     & 0.427	        \\
120   & 0.10     & 2.900		   & 47.64	    & 3.007		   & 44.48	  & 0.010	       & 0.077        & 0.851	      & 0.077	     & 0.548 \\
60    & 0.5      & 3.394                   & 29.56          & 4.357                & 18.34        & 0.0                & 0.123 & 0.834 & 0.0 & 0.975\\
60    & 0.25     & 3.807                   & 36.31          & 4.213                & 20.94        & 0.0                & 0.113        & 0.834 & 0.0 & 0.978\\ 
60    & V        &  3.685                  & 40.60          & 3.787                & 26.20        & 0.719              &  0.228        & 0.023        & 0.0       & 0.876\\

\hline
\end{tabular}	
\end{onecolumn}														       
\label{tab:tracks}
\end{table}
%


\begin{table*}

   \caption[]{Some of the most luminous stars  in the Galaxy and Magellanic Clouds. }
\renewcommand{\arraystretch}{0.7}
\begin{tabular}{lllllllll}
\hline
\hline   
  No. & Star       & Sp. type       & $\log L/L_\odot$ & $\log T_{\rm eff}$ & \md, $10^{-6}$\,\myr & $M/\ms$  & Ref.      & Comment\\[0.2cm]         
1  & HD5280A    & LBV   	& 6.5              & 4.724              &         	        &	50.   &    1     & SMC, Y/X=0.63 \\
2  & HD5280B    & WNE?           & 6.3              & 4.748  	         &         		&	28.   &    1	&  Comp. to A., X=0 \\
3  & HD5280C    & O4-6            & 6.1              & 4.665  	         &         		&	     &    1	& N enhanced      \\
4  & Sk 80      & O7If          &6.1              & 4.560               &                       &           &       2    & SMC \\   
5  & MPG355     & ON3III(f*),   & 6.98   	    & 4.690              &         		&	     &      2	  & SMC		     \\
6  & R136-040   &  O2-3.5 V      & 5.82             &   4.71      	 &         	  2.0	&	25.  &         3  & LMC\\
7  & BI 253     & O2 V((f*))	&  5.82              &  4.68	         &         	   3.5  &	43.  &         3  & \ldots \\
8  & BI 237     &  O2 V((f*))	&  5.77 	     &  4.68	         &         	   2.0  &	37.  &         3  & \ldots \\
9  & AV 435     &  O3 V((f*))	&  5.87              &  4.653	         &         	   0.5  &	48.  &         3  & SMC\\  	   
10 & AV 14      &  O5 V		&  5.85              &  4.643	         &         	   0.3  &	75.  &         3  & SMC\\
11 & LH64-16    &  ON2 III(f*)	&  5.85              &  4.74	         &         	   4.0  &	26.  &         3  & LMC \\
12 & R136-047   &  O2 III(f*)	&  5.82              &  4.71	         &         	   6.0  &	32.  &         3  & \ldots \\
13 & R136-018   &  O3 III(f*)	&  5.9               &  4.65	         &         	   2.0  &	46.  &         3  &  \ldots \\
14 & LH90:ST2-22 &  O3.5 III(f+) & 6.08		     &  4.643	         &         	   4.5  &	67.  &         3  &  \ldots \\
15 & LH101:W3-19 & O2 If*  	 & 6.34 	     &  4.643	         &         	   20.  &	193. &         3  &  \ldots \\
16 & R136-036    & O2 If*  	 & 5.7  	     &  4.633	         &         	   14.  &	31.  &         3  &  \ldots \\
17 & R136-020    & O2 If*  	 & 5.9  	     &  4.628	         &         	   23.  &	40.  &         3  &  \ldots \\
18 & Sk -67 22   & O2 If*  	 & 5.69              &  4.623            &         	   15.  &	23.  &         3  &  \ldots \\
19 & LH90:Br58   & O3If/WN6  	 & 5.9  	     &  4.613	         &         	   40.  &	40.  &         3  &  \ldots \\
20 & R136-014    & O3.5 If*	 & 5.9  	     &  4.58	         &         	   23.  &	53.  &         3  &  \ldots \\
21 & Sk -65 47   & O4 If	 & 5.97		    &   4.60	         &         	   12.  &	61.  &         3  & \\
22 & AV 75       & O5.5 I(f)	 & 6.16 	    &   4.60	         &         	   3.5  &	96.  &         3  & SMC\\
23 & AV 26       & O6 I(f) 	 & 6.14 	    &   4.58	         &         	   2.5  &	91.  &         3  &  \ldots \\
24 & HD93250     & O3 V ((f))	 & 6.01		    &  4.662             &         	   3.45 &	83.3 &         4  & \ldots \\
25 & HD66811     & O4 I(f)       & 5.90             & 4.613              &         	    8.8 &	53.9 &         4  & \ldots \\
26 & HD14947     & O5 If+	 & 5.90		    &  4.574             &         	   8.52 &	30.7 &         4  &  \ldots \\
27 & HD15558     & O5 III(f) 	& 5.93	            &	 4.613	         &         	   5.58 &	78.7 &         4  &  \ldots \\
28 & HD210839    & O6 I(n) fp	& 5.83	            &	 4.556		 &         	6.85	&	 62.2&    	4 & \ldots \\
29 & HD30614     & O9.5 Ia	& 5.83              &	 4.462           &         	6.04	&	 37.6&         4  &  \ldots \\
30 & KY Cyg      & M3-4 I	& 5.43 - 6.04               &	  3.544 	 &         		&	     &  5 & MW\\
31 & BD+60 2613  & M3 I		& 5.32 - 5.75              &	3.544		 &    	   	        &	&       5 & \ldots \\
32 & Mk51       &  O3I*f/WN7-A  & 6.23       & 4.648 & & &6 & LMC \\
33 & R139       &  O7Iafp       & 6.40       & 4.558 & & &6 & \ldots \\
34 & Mk26       &  O4 III(f)    & 6.02       & 4.647 & & &6 & \ldots \\
35 & Mk24       &  O3V          & 6.00       & 4.686 & & &6 & \ldots \\
36 & Mk14       &  O3-6V        & 5.94       & 4.646 & & &6 & \ldots \\
37 & IRC+10420  &  OH/IR        & 5.7        & 3.845 & 300-600& & 7& MW \\
38 & CygOB2 No.9 & O5f          & 6.6        & 4.650 & 12.7     & 160. & 8 & \ldots\\
39 & CygOB2 No.8A & O5.5 I(f)   & 6.18       & 4.585 & 13.5   & 118. & 8 &  \ldots  \\
40 & CygOB2 No.12 & B8 Ia       & 6.20       & 4.049 & 38.5   & 71.  & 8 & \ldots \\ 
41 & HD 33579     & A3 Ia$^+$   & 5.72       & 3.902 & 2.     & 20.-30. & 9 & LMC\\ 
42 & HD80077      & B2 Ia$^+$   & 6.4        & 4.23  &        &         & 10 & MW, (LBV?)\\
43 & HD119796      & G8 Ia$^+$   & 5.7        & 3.67 &        &         & 10 &  \ldots \\
44 & HD152234     & B05 Ia(N wk) & 5.87      & 4.41 &  2.7   &         & 11 &  \ldots  \\
45 & HD15236      & B1.5 Ia      & 6.1       & 4.26 &  6.0   &         & 11 &  \ldots  \\
46 & HD224914 ($\rho$ Cas) & F8 Ia$^+$ &5.70 & 3.86 & 54000.  &         & 11, 12 &  \ldots , $\dot{M}$ in outburst\\

     \hline									      		
\end{tabular} 


References: 1 --  \citet[][]{koenig04}, 
2 -- \citet[][]{massey89}, 
3 -- \citet[][]{massey05},
4 -- \citet[][]{repolust04},
5 -- \citet[][]{levesque05},
6 -- \citet[][]{walborn97},
7 -- \citet[][]{humphreys_irc97},
8 -- \citet[][]{waldron04},
9 -- \citet[][]{NJ_HD33579},
10 -- \citet[][]{dejager98},
11 -- \citet[][]{crowther_06},
12 -- \citet[][]{lobel_rhocas03}
\label{tab:obs}
\end{table*}


\newpage

\newpage

\end{document}